\documentclass[3p,twocolumn,fleqn]{elsarticle}
\newcommand{\bqn}{\begin{equation}}
\newcommand{\eqn}{\end{equation}}
\newcommand{\bqna}{\begin{eqnarray}}
\newcommand{\eqna}{\end{eqnarray}}
\newcommand{\bary}{\begin{array}{clcr}}
\newcommand{\eary}{\end{array}}
\usepackage{graphicx}
\usepackage{multicol}
\usepackage{color}
\usepackage{xspace}
\usepackage{enumerate}
\usepackage{amsmath}
\usepackage{color}
\usepackage{amsfonts}
\usepackage{array}
\usepackage{float}
\usepackage{bm}
\usepackage{setspace}
\begin{document}

\title
{Propagation of two short laser pulse trains in a $\Lambda$-type three-level
medium under conditions of electromagnetically induced transparency}

\author[GB,TN]{Gabriela Buica}
\ead{buica@spacescience.ro}

\author[TN]{Takashi Nakajima \corref{cor1}}
\ead{t-nakajima@iae.kyoto-u.ac.jp}

\cortext[cor1]{Corresponding author}

\address[GB]{Institute of Space Science, Bucharest-M\u{a}gurele,
 P.O. Box MG-23, R-77125, Romania}
\address[TN]{Institute of Advanced Energy, Kyoto University, Gokasho, Uji,
Kyoto 611-0011, Japan}


\begin{abstract}
We investigate the dynamics of a pair of short laser pulse trains propagating
in a medium consisting of three-level $\Lambda$-type atoms
by numerically solving the Maxwell-Schr\"{o}dinger equations for  atoms
and fields. By performing propagation calculations with different parameters,
under conditions of electromagnetically induced transparency,
we compare the propagation dynamics by a single pair
of probe and coupling laser pulses and by probe and coupling laser pulse trains.
We discuss the influence of the coupling pulse area,
number of pulses, and detunings on the probe laser propagation
 and realization of electromagnetically induced transparency  conditions,
 as well on the formation of a dark state.

\end{abstract}
%\pacs{42.50.Gy, 32.80.Qk, 42.65.Re}
\maketitle
\date{\today}

\section{Introduction}

Quantum control, sometimes referred to as coherent control, has drawn
increasing interest in recent years in many different areas of physical
sciences. Quantum control aims at manipulating the fate of quantum systems
at will by utilizing quantum interference in one way or the other
\cite{Shapiro2000,Ehlotzky}.
The optical response of an atomic medium can be modified due to
quantum interference between two different excitation pathways and
an opaque optical medium can be rendered  transparent to a probe field
by applying an intense coupling laser field at a different frequency
\cite{Harris,Marangos}.
This phenomenon has been termed electromagnetically induced transparency
(EIT), and a narrow transparency window with vanished absorption and
refractive index appears within an absorption line.
EIT was theoretically proposed by Kocharovskaya and coworkers
\cite{Olga}, and then it was experimentally demonstrated with Sr atoms
 by Harris and coworkers \cite{Boller}.
The basis of EIT resides in the \textit{coherent population trapping} (CPT)
 in three-level $\Lambda$-type atoms, which was first discovered by
Alzetta and coworkers \cite{Alzetta} in the D lines of Na atoms:
Briefly, for $\Lambda$-type atoms
interacting with a probe and a coupling laser the population is
trapped in the two lower states without excitation to the upper
(intermediate) state during the interactions.
The underlying physics of CPT is the destructive quantum interference
for the transitions from the two lower states to the common upper
(intermediate) states, where the establishment of coherence between
the two lower states by the probe and coupling lasers is the key.
Since the first experimental demonstration of EIT \cite{Boller}, various
kinds of EIT-related phenomena have been intensively studied, some of which
are lasing without inversion, nonlinear optics, sub-fs pulse generation,
atomic coherence control, slow light, giant nonlinearity, and
storage of light, etc. \cite{Fleischhauer}.

Quantum interference can be induced by lasers in many different ways.
For instance the incident laser may be a single laser pulse or a
\textit{laser pulse train}.
By using a comb laser, which means an ultrafast laser pulse train with a
high repetition rate, one can carry out the ultrahigh resolution spectroscopy.
Recently it has been shown that the \textit{accumulation effects of coherence}
by a laser pulse train play an important role for coherent control of atomic
or molecular systems \cite{Marian,Stowe}.
The use of a {\it chirped} laser pulse train is another way for coherent
control of population transfer \cite{Shapiro2007}.
Recent advances of laser technology allows us to actively control the
amplitude and phase of the laser field from pulse to pulse, and
the application of such techniques to a frequency-comb laser
irradiated to a $\Lambda$-type molecule results in a robust population
transfer \cite{Shapiro2008}.

So far most of the studies on the $\Lambda$-type system with a femtosecond
laser pulse train have been done, without taking into account the propagation
effects, either by assuming the presence of a stationary state
\cite{Felinto2003} or a weak laser field when the lowest order
perturbation theory is applicable \cite{Felinto2004}.
As for the work with propagation effects taken into account,
a closed $\Lambda$-type system with a degeneracy in the two lower states
has been theoretically investigated with a femtosecond laser pulse train
\cite{Soares2009}, when the \textit{single} laser pulse train acts
as both probe and coupling laser pulses due to the degeneracy of the system
under consideration.  At first glance such a degenerate $\Lambda$-type
system looks very similar to a non-degenerate $\Lambda$-type system
which is most commonly studied in the context of EIT.
There is, however, an essential difference between them:
The initial state of the former is a {\it mixed state}
with 50:50 populations in the two degenerate lower states without coherence,
while for the latter the initial state is a {\it pure state} and only a single
state is occupied before the interactions with laser pulses.
As a result, in terms of the single atom response, i.e., at the entrance to
the medium, a dark state is not formed in the degenerate $\Lambda$-type
system by the very first pulse, while is formed in the non-degenerate
$\Lambda$-type system.
As a natural consequence, we expect that the following propagation dynamics
would be essentially different.

In this paper we numerically investigate the propagation dynamics of
\textit{a pair} of short laser pulse trains in a non-degenerate
$\Lambda$-type atomic medium under the EIT conditions when the coupling
field is stronger than the probe field.
In principle such {\it two-colour} laser pulse trains can be produced
from a single laser pulse train by the optical parametric amplification
technique, etc.
The pulses we assume in this paper are short, and the time interval
between two successive pulses is also short compared with a lifetime
of the excited state, as a result of which the spontaneous decay
from the excited state will not be completed before the next pulses arrive.
 Unlike the case of the degenerate $\Lambda$-type system, however,
this fact would play a minor role for the non-degenerate $\Lambda$-type
system, since the first pulse produces a nearly perfect dark state
under the conditions of EIT, at least at the entrance to the medium,
and hence the population in the excited state is almost zero.
The paper is organized as follows.
In Section  \ref{mm}  we introduce the theoretical model.
To describe the propagation dynamics we utilize the Maxwell-Schr\"{o}dinger
equations for  atoms and fields and numerically solve them on a grid,
assuming a one-dimensional propagation.
In Section \ref{nr} we present and discuss representative numerical results for the
 interaction of a pair of probe and coupling laser as well as
 probe and  coupling laser pulse trains with a three-level $\Lambda$-type atom.
The question we address in this paper is whether and how much the propagation
dynamics of a pair of laser pulse trains in a non-degenerate $\Lambda$-type
atomic medium is different from that of a single pair of pulses under
the EIT conditions.
We examine the effect of the laser parameters such as coupling pulse area,
number of individual pulses, and laser detunings on the temporal and
 spatial propagation dynamics of the probe laser pulse train.
Finally, concluding remarks are given in Section \ref{su}.
Atomic system of units are used throughout the present work unless otherwise
stated.

\section{Theoretical model}
\label{mm}

\subsection{Laser pulse trains}

In Fig. 1 we show the level scheme of a $\Lambda$-type atom interacting with
a pair of short laser pulse trains:
the state $| 1 \rangle$ is initially occupied, while the states $| 2 \rangle$ and
$| 3 \rangle$ are initially unoccupied, and the probe (coupling) laser field
resonantly couples the states $| 1 \rangle$ and $| 2 \rangle$
($| 2 \rangle$ and $| 3 \rangle$), respectively.
The transition between the states $| 1 \rangle$ and $| 3 \rangle$ is
dipole-forbidden.
The total electric field vector is written as

\bqn
\textbf{E}(z,t) =\textbf{E}_p(z,t) +\textbf{E}_c(z,t) \;,
\label{field}
\eqn

\noindent
where $\textbf{E}_p(z,t)$ and $\textbf{E}_c(z,t)$ are the
probe and coupling electric field vectors that are copropagating along the $z $ axis.
 Since we assume that they are
in the form of pulse trains with linear polarizations which are parallel
to each other, the respective fields are written as

%\bqn
\begin{flalign}
\textbf{E}_{\alpha}(z,t)& =
\mathcal{E}_{\alpha 0} \;\bm {e}\; \exp{[i(\omega_{\alpha} t- k_{0 \alpha } z ) ]}\nonumber \\
&\times \sum_{n=0}^{N_{max}-1} f_{\alpha} ( z,t - n T ) e^{i  n\phi} + c.c.
\;,
\label{field_p}
\end{flalign}
%\eqn

\noindent
where $\mathcal{E}_{\alpha 0} $ $($with $\alpha=p \mbox{ or } c$ hereafter$)$
are the peak field amplitudes of the probe and coupling laser fields,
respectively, and $\bm{e}$ is the polarization vector which is assumed
to be identical for both lasers.
$f_{\alpha}(z,t)$ is the slowly varying envelope, $\omega_\alpha$ is
the photon energy, and $k_{0 \alpha }$ is the wavenumber in vacuum for the
respective laser fields.
$T$ and $N_{max}$ represent the time interval between two successive pulses
and the number of pulses in each pulse train, respectively, which are assumed
to be identical for both lasers, and $\phi$ is a phase shift between
two successive pulses.

By taking the Fourier transform of the electric field Eq. (\ref{field_p})
 with respect to time as
$$
\widetilde E_\alpha(z,\omega) =
\int_{-\infty} ^{+\infty} E_\alpha(z,t) \exp{(-i\omega t)} dt,
\nonumber
$$
\noindent
where the symbol $\sim$  denotes a  Fourier transform,
we obtain  by applying the Poisson formula  for  $N_{max} \to \infty $,
after a simple algebra,

\bqn
\widetilde E_\alpha(z,\omega) = \widetilde E_{\alpha0}(z,\omega - \omega_{\alpha})
\sum_{m=-\infty} ^{+\infty} \delta( \omega -\omega_{\alpha m}),
\label{field_p_spec}
\eqn

\noindent
with  $\widetilde E_{\alpha0}(z,\omega - \omega_{\alpha})
= 2\pi \; \mathcal{E}_{\alpha 0} \widetilde f_{\alpha}(\omega - \omega_{\alpha}) /T $
and $\omega_{\alpha m} = \omega_{\alpha}  + \phi / T + 2 \pi m / T$.
The laser pulse train spectrum defined by Eq. (\ref{field_p_spec}) forms a frequency comb
with $m$ laser modes centered at $\omega_{\alpha} + \phi / T$,
separated by $2\pi/T$, and  with a peak amplitude
$\widetilde E_{\alpha0}(z,\omega - \omega_{\alpha})$.
Therefore the photon energy of the $m^{th}$ mode for each pulse train is
given by $ (\omega_{\alpha} + \phi / T) \pm 2 \pi m / T$.
In this paper we assume that the phase shift, $\phi$, is zero.
As for the slowly varying envelope at the entrance to the medium,
$f_{\alpha}(z=0,t)$, we assume that it is represented by a Gaussian
function, i.e.,

\bqn
f_{\alpha}(z=0,t) = e^{ -\pi \left( {t}/{\tau_{0}} \right)^2 }
 \; ,
\label{field_gauss}
\eqn

\noindent
where $\tau_{0}$  is the pulse temporal width of a single laser pulse,
which is assumed to be identical for both probe and coupling lasers.

\subsection{Maxwell-Schr\"{o}dinger equations}\label{am_eq}

As long as the pulse duration is not very short, say, if it is longer than
100 fs, we may derive the equations for atoms and fields,
which are often called Maxwell-Schr\"{o}dinger equations, after introducing
the rotating-wave approximation and slowly varying envelope
approximation (SVEA).
In SVEA the envelope of the electric field varies slowly in time and space
compared to the optical period and the wavelength of the field, and
therefore the second order derivative terms are negligible, i.e.,
$ |\partial_z^2 f_{\alpha}(z,t)| \ll k_{0 \alpha} |\partial_z f_{\alpha}(z,t)|$ and
$ |\partial_t^2 f_{\alpha}(z,t)| \ll  \omega_{\alpha} |\partial_t f_{\alpha}(z,t)|$,
where $\partial_{z}$ and $\partial_{t}$ represent the derivatives relative
to the variables $z$ and $t$.

As for the response of the single atom to the laser fields,
we can follow the standard procedure and easily derive the time-dependent
Schr\"{o}dinger equations in terms of the probability amplitudes of the
relevant states described in Fig. 1. Since we couple those equations
with the Maxwell equations in what follows, it is more convenient to
introduce a moving frame by replacing the space and time variables in the
laboratory frame, $z$ and $t$, by those in the moving frame, $\zeta$ and
$\tau$, through the relations of $\zeta=z$ and $\tau=t-z/c$, where $c$
is the speed of light.
Finally, we obtain the following probability amplitude equations

%\bqna
\begin{flalign}
 \frac{\partial}{\partial \tau} {c}_{1}(\zeta,\tau) &=
\; \frac{i}{2} {\Omega}_p (\zeta,\tau) c_{2}(\zeta,\tau) ,
\label{amp1} \\
 \frac{\partial}{\partial \tau} {c}_{2}(\zeta,\tau) &=
-(i \delta_p + \gamma) c_{2}(\zeta,\tau)
\label{amp2} \\
&+ \; \frac{i}{2} {\Omega}_p (\zeta,\tau) c_{1}(\zeta,\tau)
+ \; \frac{i}{2} {\Omega}_c (\zeta,\tau) c_{3}(\zeta,\tau) ,
\nonumber \\
  \frac{\partial}{\partial \tau} {c}_{3}(\zeta,\tau)& =
-i(\delta_p - \delta_c ) c_{3}(\zeta,\tau)
\nonumber \\
&+ \; \frac{i}{2} {\Omega}_c (\zeta,\tau) c_{2}(\zeta,\tau) ,
\label{amp3}
\end{flalign}
%\eqna

\noindent
where
${c}_{k}(\zeta,\tau)$ ($k=1,2$, and $3$) is the slowly varying probability
amplitude of state $| k \rangle$ with the initial conditions of
${c}_{1}(\zeta,\tau=-\infty)= 1$ and
${c}_{k}(\zeta,\tau=-\infty) = 0$ ($k=2$, $3$),
$\gamma$ is the spontaneous decay rate of state $| 2 \rangle$, and
$\delta_{p}$ and $\delta_{c}$ are the detunings of the probe and coupling
lasers, respectively.
$\Omega_{p} (\zeta,\tau)$ and $\Omega_{c} (\zeta,\tau)$  are the one-photon
Rabi frequency due to the probe and coupling laser pulse trains, and are defined
as

\begin{flalign}
{\Omega}_{p} (\zeta,\tau) &= {\Omega}_{p0} \sum_{n=0}^{N_{max}-1} f_{p} ( \zeta,\tau - n T ),
\label{op}\\
{\Omega}_{c} (\zeta,\tau) &= {\Omega}_{c0} \sum_{n=0}^{N_{max}-1} f_{c} ( \zeta,\tau - n T ),
\label{oc}
\end{flalign}
\noindent
with the peak value of ${\Omega}_{p0}= d_{12} {\mathcal{E}}_{p0}$
(${\Omega}_{c0}= d_{23} {\mathcal{E}}_{c0}$), where $d_{12}$ ($d_{23}$)
is the one-photon dipole moment for the transition between states
$| 1 \rangle$ and $| 2 \rangle$ ($| 2 \rangle$ and $| 3 \rangle$),
respectively.

As for the response of the laser fields to the atomic medium
we obtain the following Maxwell equations

\begin{flalign}
\frac{\partial}{\partial \zeta} \Omega_p(\zeta,\tau)&
= -2 i \mu_p   c_2 ^*(\zeta,\tau)c_1(\zeta,\tau) ,
\label{max1} \\
 \frac{\partial}{\partial \zeta} \Omega_c(\zeta,\tau)&
= -2 i \mu_c  c_2^* (\zeta,\tau)c_3 (\zeta,\tau) ,
\label{max2}
\end{flalign}

\noindent
where $\mu_p$ ($\mu_c$) is the medium propagation coefficient for the
probe (coupling) laser, which is defined as
 $\mu_{p} = {N_d \omega_{p}|d_{12}|^2 }/{ 2 \varepsilon_0 c}$
($\mu_{c} = {N_d \omega_{c}|d_{23}|^2 }/{ 2 \varepsilon_0 c}$),
where $N_d$ is the atomic density of the medium and $\varepsilon_0$ is the
vacuum permittivity.
The initial conditions for Eqs. (\ref{max1})-(\ref{max2})
 are given by Eqs. (\ref{field_gauss}),  (\ref{op}) and (\ref{oc}).

Now, we solve the coupled differential
Eqs. (\ref{amp1})-(\ref{amp3}),  (\ref{max1}) and (\ref{max2})
for the atoms and laser fields with arbitrary temporal shapes and field
strengths.
This is a complicated mathematical problem and there is no analytical
solution for the general case. Therefore we resort to the numerical method
based on the Crank-Nicholson algorithm which has a second order accuracy
in both time and space. The computer code we have developed for this problem
is an extension of our previous work for the propagation of phase-controlled
two-colour lasers in a two-level medium \cite{Nakajima2001}.

\subsection{Maxwell-Schr\"{o}dinger equations in the dressed state basis}
\label{am_eqdb}

A more convenient way to study the Maxwell-Schr\"{o}dinger equations
is to use the  dressed state basis \cite{Harris1995},
by taking the advantage of the \textit{dark} and \textit{bright}
states, $|D\rangle$ and $|B\rangle$,  defined as coherent asymmetric
 and symmetric superposition  of the two lower bare states,
$|1 \rangle$ and $|3 \rangle$,
$ |D\rangle =(\Omega_{c0} |1 \rangle  - \Omega_{p0}|3 \rangle)/\Omega_{B0}$
and $ |B\rangle =(\Omega_{p0} |1 \rangle  + \Omega_{c0}|3 \rangle)/\Omega_{B0}$,
where $\Omega_{B0}=\sqrt{\Omega_{p0}^2+\Omega_{c0}^2 }$ represents
the bright Rabi frequency.
The \textit{dark} and \textit{bright} probability amplitudes,
$ c_D $ and $ c_B $, are now calculated in terms of $ c_1 $ and $ c_3 $,

%\bqna
\begin{flalign}
& c_D(\zeta,\tau)= \frac{1}{\Omega_{B0}}
[ \Omega_{c0} c_1(\zeta,\tau)     - \Omega_{p0}c_3(\zeta,\tau) ] ,
\label{darkstate}\\
& c_B(\zeta,\tau)= \frac{1}{\Omega_{B0}}
[ \Omega_{p0} c_1(\zeta,\tau)  + \Omega_{c0}c_3(\zeta,\tau) ]  .
\label{brightstate}
\end{flalign}
%\eqna

\noindent
Similarly,  we define new field variables \cite{Harris1995},  $\Omega_D(\zeta,\tau)$
 and  $ \Omega_B(\zeta,\tau)$,  the \textit{dark} and
\textit{bright} fields, calculated in terms of $ \Omega_p$ and $ \Omega_c$,  as

%\bqna
\begin{flalign}
\Omega_D(\zeta,\tau) &= \frac{1}{\Omega_{B0}}
[  \Omega_{c0} \Omega_p(\zeta,\tau)   -  \Omega_{p0} \Omega_c(\zeta,\tau) ] ,
\label{darkfield} \\
\Omega_B(\zeta,\tau) &=\frac{1}{\Omega_{B0}}
[ \Omega_{p0} \Omega_{p}(\zeta,\tau)     + \Omega_{c0} \Omega_{c}(\zeta,\tau)] .
\label{brightfield}
\end{flalign}
%\eqna

\noindent
First, by taking the  derivative of Eqs. (\ref{darkstate}) and (\ref{brightstate})
 with respect to time, $\tau$,
and using the  probability amplitude equations  (\ref{amp1})-(\ref{amp3})
and dressed fields' definitions  (\ref{darkfield}) and (\ref{brightfield})
we obtain the following set of probability amplitude
equations in the new basis of the dressed states:

%\bqna
\begin{flalign}
& \frac{\partial}{\partial \tau} {c}_{D}(\zeta,\tau) =
 \frac{i}{2} {\Omega}_{D} (\zeta,\tau) c_{2}(\zeta,\tau)
 \label{amp1d} \\
& -i(\delta_p - \delta_c )
 \frac{\Omega_{p0}} {\Omega_{B0}^2}
\left[\Omega_{c0}c_B(\zeta,\tau) +\Omega_{p0}c_D(\zeta,\tau) \right]
, \nonumber \\
& \frac{\partial}{\partial \tau} {c}_{2}(\zeta,\tau)  =
-(i \delta_p + \gamma) c_{2}(\zeta,\tau) \label{amp2d}\\
&
+ \; \frac{i}{2} {\Omega}_{B} (\zeta,\tau) c_{B}(\zeta,\tau)
+ \; \frac{i}{2} {\Omega}_{D} (\zeta,\tau) c_{D}(\zeta,\tau) ,
\nonumber \\
&  \frac{\partial}{\partial \tau} {c}_{B}(\zeta,\tau) =
 \frac{i}{2} {\Omega}_{B} (\zeta,\tau) c_{2}(\zeta,\tau)
\label{amp3d} \\
& -i(\delta_p - \delta_c )   \frac{\Omega_{c0}} {\Omega_{B0}^2}
\left[\Omega_{p0}c_B(\zeta,\tau) +\Omega_{c0}c_D(\zeta,\tau) \right]
,\nonumber
\end{flalign}
%\eqna
with the initial conditions
${c}_{D}(\zeta,\tau=-\infty)= \Omega_{c0}/{\Omega}_{B0}$,
${c}_{2}(\zeta,\tau=-\infty) = 0 $, and
${c}_{B}(\zeta,\tau=-\infty)= \Omega_{p0}/{\Omega}_{B0}$,
as the $\Lambda$-type atom is initially prepared in the ground state.
Second, by taking the  derivative  of
Eqs. (\ref{darkfield}) and (\ref{brightfield}) with respect to $\zeta $  and
 using Eqs. (\ref{max1})-(\ref{brightstate})
we get after a simple algebra the following Maxwell
equations in the dressed state basis:

%\bqna
\begin{flalign}
  \frac{\partial}{\partial \zeta} \Omega_D(\zeta,\tau) &=
 -2 i   c_2 ^*(\zeta,\tau)
\label{max1d} \\& \times
\left[ \mu_{B1}  c_D(\zeta,\tau)  + \mu_{D}  c_B(\zeta,\tau) \right] ,
\nonumber\\
  \frac{\partial}{\partial \zeta} \Omega_B(\zeta,\tau) &=
 -2 i   c_2 ^*(\zeta,\tau)   \label{max2d} \\ & \times
\left[  \mu_{D}  c_D(\zeta,\tau)  +  \mu_{B2}  c_B(\zeta,\tau) \right] ,
 \nonumber
\end{flalign}
%\eqna

\noindent
where the new propagation coefficients $\mu_{B1}, \mu_{B2} $ and $ \mu_D$ are defined as
$\mu_{B1} = (\Omega_{c0}^2\mu_p    + \Omega_{p0}^2\mu_c) /{\Omega_{B0}^2}$,
$\mu_{B2} = (\Omega_{p0}^2\mu_p    + \Omega_{c0}^2\mu_c )/{\Omega_{B0}^2} $,
 and $ \mu_D= (\mu_p - \mu_c )\Omega_{p0} \Omega_{c0} /{\Omega_{B0}^2}$, respectively.
For initially matched probe and coupling pulses at the entrance to the medium,
$f_p(\zeta=0,\tau)=f_c(\zeta=0,\tau)$,
the initial conditions satisfied by the dressed fields are evaluated
from Eqs. (\ref{darkfield}) and (\ref{brightfield}) as
${ \Omega}_{D}(\zeta=0,\tau)= 0$ and
${ \Omega}_{B}(\zeta=0,\tau)= \Omega_{B0} \sum_{n} f_{p} ( \zeta=0,\tau - n T )$.

Finally, because we are interested in a weak interaction of the probe field
with the atom, $\Omega_{p0} \ll \Omega_{c0}$,
important simplifications of the probability amplitude Eqs.
 (\ref{amp1d})-(\ref{amp3d}) occur as follows,

%\bqna
\begin{flalign}
 \frac{\partial}{\partial \tau} {c}_{D}(\zeta,\tau) &\simeq 0,
\label{amp1ds} \\
   \frac{\partial}{\partial \tau} {c}_{2}(\zeta,\tau)  &\simeq
-(i \delta_p + \gamma) c_{ 2 }(\zeta,\tau) \label{amp2ds} \\
&+ \; \frac{i}{2} {\Omega}_{B} (\zeta,\tau) c_{B}(\zeta,\tau),
 \nonumber\\
  \frac{\partial}{\partial \tau} {c}_{B}(\zeta,\tau) &\simeq
-i(\delta_p - \delta_c ) c_{D}(\zeta,\tau) \label{amp3ds} \\
&+\; \frac{i}{2} {\Omega}_{B} (\zeta,\tau) c_{2}(\zeta,\tau),
  \nonumber
\end{flalign}
%\eqna

\noindent
with the initial conditions
${c}_{D}(\zeta,\tau=-\infty) \simeq 1$,
${c}_{2}(\zeta,\tau=-\infty) = 0 $, and
${c}_{B}(\zeta,\tau=-\infty) \simeq 0 $,
 showing that  the $\Lambda$-type atom is
initially in a dark state $|D\rangle$.
Obviously, Eqs.  (\ref{amp1ds})-(\ref{amp3ds})
 correspond to a $\Lambda$-type atom
 where the dressed state $|D\rangle$ is practically decoupled from the
 two fields and the dark field is,
 accordingly to its definition (\ref{darkfield}),
 $\Omega_D(\zeta,\tau) \simeq 0$,
reducing thus the problem of a three- to a two-level atom.
Hence,  for initially matched pulses in the limit of  weak probe field,
 $\Omega_{p0} \ll \Omega_{c0}$,
  we obtain that the atomic system reaches a dark state where
$\Omega_D(\zeta,\tau) \simeq 0$  and
 the temporal  profile of probe and coupling laser pulses does not change
at any optical depth,
$f_p(\zeta,\tau)=f_c(\zeta,\tau) = f _p(\zeta=0,\tau)$, and therefore
the medium becomes transparent to both lasers \cite{Harris1995}.

\noindent
Similar simplifications exist for the Maxwell Eqs.
(\ref{max1d}) and  (\ref{max2d}) in the limit of weak probe field,
where the propagation coefficients simplify as
 $\mu_{B1} \simeq \mu_{p}$, $\mu_{B2} \simeq \mu_{c}$, and $\mu_{D} \simeq 0$.
Substituting the above simplified propagation coefficients in the Maxwell
Eqs. (\ref{max1d}) and  (\ref{max2d}) we obtain

%\bqna
\begin{flalign}
 &\frac{\partial}{\partial \zeta} \Omega_D(\zeta,\tau)
\simeq -2 i  \mu_{p} c_2 ^*(\zeta,\tau)    c_D(\zeta,\tau) ,
\label{max1ds}\\
 &\frac{\partial}{\partial \zeta} \Omega_B(\zeta,\tau)
\simeq  -2 i  \mu_{c} c_2 ^*(\zeta,\tau)    c_B(\zeta,\tau)   .
\label{max2ds}
\end{flalign}
%\eqna

Next, because the atom interacts with two laser pulse trains, whose spectra
consist of combs with different frequencies, is useful to consider
the  above Maxwell-Schr\"{o}dinger equations (\ref{amp1ds})-(\ref{max2ds})
in the spectral domain where the dependence on the frequency comb modes
 is emphasized.
By taking the  Fourier transform of Eqs. (\ref{amp1ds})-(\ref{amp3ds})
with respect to $\tau$ and using the  Fourier transform of the probability
 amplitude  $c_k(\zeta,t)$ ($k=B,D$ and $2$) defined as
\bqn
\widetilde c_k(\zeta,\omega) = \int_{-\infty} ^{+\infty}
c_k(\zeta,\tau') e^{-i\omega \tau'} d \tau',
\eqn

\noindent
 we can write, after a straightforward algebra, the set of probability
amplitude equations  in the spectral domain as follows,

%\bqna
\begin{flalign}
 \widetilde{c}_{D}(\zeta,\omega) &\simeq 0,
\label{amp1dss} \\
  \widetilde{c}_{2}(\zeta,\omega) &\simeq
\frac{1}{2( \omega +\delta_p - i \gamma) }\label{amp2dss} \\
&\times
\sum_{m=-\infty} ^{+\infty}
\widetilde \Omega_{B0}(\zeta,m\omega_{r})
\widetilde{c}_{B}(\zeta,\omega - m\omega_{r}),
\nonumber \\
 \widetilde{c}_{B}(\zeta,\omega) &\simeq
\frac{1}{2( \omega +\delta_p - \delta_c) } \label{amp3dss} \\
&\times
\sum_{m=-\infty} ^{+\infty}
\widetilde \Omega_{B0}(\zeta,m\omega_{r})
\widetilde{c}_{2}(\zeta,\omega - m\omega_{r}),
\nonumber
\end{flalign}
%\eqna

\noindent
where we employ the Fourier transform of the bright Rabi
field, $\widetilde \Omega_B(\zeta,\omega)$, given by

%\bqna
\begin{flalign*}
\widetilde \Omega_B(\zeta,\omega) ={}&
\int_{-\infty} ^{+\infty} \Omega_B(\zeta,\tau') e^{-i\omega \tau'} d \tau'
\nonumber\\
= {} & 2 \pi \;\widetilde \Omega_{B 0}(\zeta,\omega)
\sum_{m=-\infty} ^{+\infty} \delta( \omega - m\omega_{r} ),\nonumber
%\label{rabi_p_spec}
\end{flalign*}
%\eqna
with
\[\widetilde \Omega_{B0}(\zeta,\omega) =
[ \Omega_{p0} \widetilde \Omega_{p0}(\zeta,\omega)
+ \Omega_{c0} \widetilde \Omega_{c0}(\zeta,\omega) ]/\Omega_{B0},\]
\[ \widetilde {\Omega}_{p0}(\zeta,\omega)
= d_{12}\mathcal{E}_{p0} \widetilde{f}_p(\zeta,\omega)/T ,\]
\[\widetilde {\Omega}_{c0}(\zeta,\omega)
= d_{23} \mathcal{E}_{c0} \widetilde{f}_c(\zeta,\omega)/T,\]
where $\omega_{r} =  2 \pi  / T$ represents the repetition angular frequency.
For the EIT parameters we employ, is quite visible from the amplitude probabilities
Eqs. (\ref{amp1dss})-(\ref{amp3dss}),
that the upper and bright state populations,
$|\widetilde c_k(\zeta,\omega)|^2 $ ($k=2, B$), are periodic
functions with the same periodicity as the laser frequency comb $\omega_{r}$.

By taking now the  Fourier transform of Eqs. (\ref{max1ds}) and  (\ref{max2ds})
and using Eq. (\ref{amp1dss}) the Maxwell equations in the spectral domain reads as

%\bqna
\begin{flalign}
 &\frac{\partial}{\partial \zeta} \widetilde \Omega_D(\zeta,\omega)
\simeq0,
\label{max1dspec}\\
 & \frac{\partial}{\partial \zeta} \widetilde \Omega_B(\zeta,\omega)
\simeq  - i \frac{ \mu_{c} }{ \pi} \;
\widetilde {c_2^*} (\zeta,\omega) \otimes  \widetilde c_B (\zeta,\omega) ,
\label{max2dspec}
\end{flalign}
%\eqna

\noindent
where the symbol $\otimes $ denotes the convolution  operator which is defined as
$h (\omega) \otimes g (\omega) = \int_{-\infty} ^{+\infty} h(\omega-\omega') g(\omega') d\omega'$.
In the limit of  weak probe field for initially matched pulses,
similar to the findings in the temporal domain,
 is clear from Eq. (\ref{max1dspec})
 that the atomic system reaches a dark state where
$\widetilde \Omega_D(\zeta,\omega) \simeq 0$
and the probe and coupling laser pulses
achieve identical  spectral profiles
 $\widetilde{f}_p(\zeta,\omega) =\widetilde{f}_c(\zeta,\omega)  $.

\section{Numerical Results and Discussions}
\label{nr}

In this section we present representative results for the
propagation of the short probe and coupling laser pulse trains in
a $\Lambda$-type atomic medium by numerically solving
Eqs. (\ref{amp1})-(\ref{max2}).
We consider realistic values for the atomic and lasers parameters and
 represent the propagation length in units of $\mu_p \zeta$
 which is the so-called optical depth \cite{Eberly1995}.
The population of state $| k \rangle$ ($k=1, 2$, and $3$)
at any time and space, $(\zeta, \tau)$, is calculated as
$P_{k}(\zeta , \tau) =|c_{k}(\zeta , \tau)|^2$.
Atomic coherence between states $| 1 \rangle$ and $| 2 \rangle$ is
defined by $  c_1(\zeta,\tau) c_2^*(\zeta,\tau)$, where the real part,
Re$[  c_1(\zeta,\tau) c_2^*(\zeta,\tau)]$, stands for the refractive index
and the imaginary part,  -Im$[  c_1(\zeta,\tau) c_2^*(\zeta,\tau)]$,
stands for the absorption coefficient, respectively.
The dark and bright state populations are calculated as
$P_D( \zeta, \tau)=|c_D(\zeta, \tau)|^2$ and
$P_B(\zeta, \tau)=|c_B(\zeta, \tau)|^2$.
In what follows we present the propagation of a  single pair of resonant probe
and coupling laser pulses, next we discuss the propagation
of resonant probe and coupling laser pulse trains,
and finally we investigate the influence of symmetric and asymmetric
detunings on the propagation of laser pulse trains.

\subsection{Propagation of a single pair of resonant probe and coupling laser pulses}

To start with, we investigate the propagation of a {\it single} pair of probe
and coupling laser pulses, i.e., $N_{max}=1$ in Eq. (\ref{field_p}),
 with different pulse durations and laser intensities.
Representative results for the temporal shape of the probe pulse
$\Omega_p(\zeta,\tau)$ at different optical depths
$\mu_p \zeta= 0$, $3$, $6$, and $9$ ps$^{-1}$ are shown in
Figs. {\ref{fig2}(a)-(e) where the pulse durations of both probe
and coupling laser pulses are $100$ fs, $1$ ps, $10$ ps, $100$ ps, and
$1$ ns, respectively.
The parameters we have chosen for Fig. {\ref{fig2}}
are $\Omega_{p0} =0.04 $ THz and $\Omega_{c0} = 1$ THz for the Rabi frequencies,
 $\tau_0=1$ ps for the pulse duration,  $\delta_p=\delta_c=0$  for detunings,
and the spontaneous decay rate from the upper state is $\gamma = 70$ MHz.
Similar results, but at higher coupling laser intensities, are shown in
Figs. {\ref{fig2}(f)-(j) with $\Omega_{c0} = 10$ THz.
We note in Fig. {\ref{fig2} the quite different spatio-temporal changes
of the probe pulse with short pulse durations compared to those with
long pulse durations for which most of the EIT studies have been carried out.
As we see in Figs. {\ref{fig2}}(a)-(c) and   Figs. {\ref{fig2}}(f)-(g)
 for small and moderate coupling pulse areas
($\Omega_{c0} \tau_{0} <10$) the probe laser pulse is significantly
distorted during the propagation, and each pulse breaks up into several
sub-pulses with positive and negative amplitudes
\cite{Nakajima2001,Crisp1970,Rothenberg}.
The number of modulations at the trailing edge of the pulse increases
as the propagation distance (or optical depth) increases.
We note that, if the pulse area is small and the pulse duration is
shorter than the lifetime of the upper state, the propagation of the laser
pulse does not obey the exponential absorption  given by the Beer's law
(we should recall that the Beer's law is valid for constant laser fields) \cite{Harris1993}.
Therefore the right wing (or trailing edge) of laser pulses can propagate
for longer distances \cite{Crisp1970}.
When the pulse durations are as short as sub-ps or ps  and
the pulse areas, in particular the pulse areas of the coupling laser pulses,
are also small, there is no EIT effect.
In such cases we simply observe the modulations at the trailing edge of
the probe pulse \cite{Nakajima2001,Crisp1970}, as shown in
Figs. {\ref{fig2}(a)} and (b).
As the pulse durations become longer and accordingly the pulse areas become
larger, EIT is established with the so-called preparation loss at the
leading edge of the pulse \cite{Harris1995}, as shown in
Figs. {\ref{fig2}(d)} and (e).
If we increase the peak intensities of the coupling laser pulse
by ten times, $\Omega_{c0} =10$ THz, (right column of Fig. {\ref{fig2}}) we observe
similar dynamics but at shorter pulse durations.

For a better understanding of the break-up process of the small area pulse,
 we plot in Figs. {\ref{fig3}}(a)-(c) and Figs. {\ref{fig3}}(d)-(f)
the temporal shape of the probe pulse, $\Omega_p(\zeta,\tau)$,
and the population in the ground state, $P_{1}(\zeta,\tau)$, respectively,
at two different optical depths $\mu_p \zeta =0$ and $1$ ps$^{-1}$
for the case of a single laser pulse propagating in a \textit{two-level medium},
i.e., $N_{max}=1$ and $\Omega_{c0}=0$.
The parameters we choose for Fig. {\ref{fig3}} are $\Omega_{p0} =0.04 $ THz,
$\tau_0=5$ ps, and $\delta_p=0 $ with three different spontaneous decay rates of
$\gamma = 700 $ GHz ($\gamma^{-1} = 1.4$ ps),
$70$ GHz ($\gamma^{-1} = 14$ ps), and $7$ GHz ($\gamma^{-1} = 140$ ps)
for Figs. {\ref{fig3}(a) and (d), Figs. {\ref{fig3}(b) and (e), and
Figs. {\ref{fig3}(c) and (f), respectively.
It is well known that in the weak field regime, when the Rabi frequency
is smaller than the atomic linewidth,
$\Omega_{p0} < \gamma$ shown in Fig. {\ref{fig3}}(a),
 the laser pulse is damped and strongly absorbed,
while in the strong field regime,
$\Omega_{p0} > \gamma$ in Fig. {\ref{fig3}}(c),
the laser pulse oscillates and reshapes  due to the  Rabi flopping
 between the ground and excited state populations.
Since we are using laser pulses we should also take into consideration
the pulse duration \cite{Crisp1970}.
As we see in Fig. {\ref{fig3}}(a), the pulse is absorbed by the medium
if the lifetime of state $| 2 \rangle$ is short compared with the
pulse duration ($\gamma \tau_0 =3.5$).
If the lifetime of state  $| 2 \rangle$ is longer than the
pulse duration ($\gamma \tau_0 =0.035$), the trailing edge of the pulse modulates with positive
and negative amplitudes, as shown in  Figs. {\ref{fig3}}(b) and (c).
  The modulations of the trailing edge of the pulse
are spread over for a time scale of the order of
$\gamma^{-1}$ \cite{Harris1995}.

\subsection{Propagation of probe and coupling laser pulse trains with zero detunings}

Next, we study the propagation of the probe and coupling
\textit{laser pulse trains}, i.e., $N_{max}>1$ in Eq. (\ref{field_p}).
In Fig. {\ref{fig4}} we plot the spatio-temporal change of the probe laser
field $\Omega_p(\zeta,\tau)$, at different optical depths
$\mu_p \zeta = 0$, $3$, $6$, and $9$ ps$^{-1}$, for the first, $40^{th}$, and
$120^{th}$ pulses in the probe pulse train.
The probe and coupling laser pulse trains are initially matched with
identical pulse envelopes at the entrance to the medium.
The parameters we employ for Fig. {\ref{fig4}} are
$\Omega_{p0} = 0.04$ THz, $\Omega_{c0} =0.5$ THz, $\delta_p=\delta_c=0$,
$\tau_0=1$ ps and $T =10$ ns.
The frequency separation between two successive comb teeth is $0.2 \pi$ GHz
and the spontaneous decay rate from the upper state is $\gamma = 70$ MHz,
i.e., 14 ns lifetime.
As mentioned before the pulse duration of each individual pulse, $\tau_0$,
is much shorter than the lifetime of the upper state and there
 is no enough time for a complete decay of the upper excited state $|2\rangle$
before the next pulse in the train arrives.
Under these conditions, the area of each
of the individual probe laser pulse is small while that of the
coupling laser pulse is moderate, i.e.,
$\Omega_{p0}\tau_0 =0.04$ and $\Omega_{c0}\tau_0 =0.5$.
What we learn from Fig. {\ref{fig4}} is that, although EIT is not yet
established when the first probe pulse goes through the medium, coherence
is slowly accumulated in the medium as more probe and coupling pulses
interact with the medium, and when the 120$^{th}$ probe pulse enters
the medium we can see the some signature of EIT.
We note that our numerical results for atomic populations and probe laser
absorption at the entrance to the medium (not shown here) agree well
with the results presented in  \cite{Soares2010} in terms of a single
atom response under the presence of a probe laser train pulse and a
continuous-wave coupling laser.

What we glimpse in  Fig. {\ref{fig4}}  is indeed an EIT effect by the pair of
probe and coupling laser pulse trains but with different degrees of transparencies.
This interpretation can be verified by looking at the populations of the
\textit{dark} and \textit{bright} states.
In Fig. {\ref{fig5}} we show the variations of the dark and bright state
populations $P_D(\zeta,\tau)$ and $P_B(\zeta,\tau)$ as a function of
time at different optical depths.
As expected, the dark state population induced by the very first pulse
is nearly unity at the entrance to the medium.
This is quite in contrast to the case of a degenerate $\Lambda$-type system
in which coherence between the two lower states induced by the very first pulse
 is negligibly small, for example, as shown in the red side panel of the upper left figure
of  Fig. 6 in  \cite{Soares2009}.
Since we find out that the single atom response (at optical depth = 0) by the
first pulse is clearly different for the non-degenerate and degenerate
$\Lambda$-type systems, we expect that the following propagation dynamics
should be quite different as well.
Now, back to Fig. {\ref{fig5}} in this paper, we notice that the dark state
population becomes slightly smaller as the optical depth becomes larger
for a fixed number of irradiated pulses, say, 50, indicating that the
transparency slightly deteriorates for larger optical depths, at least
up to  $\mu_p \zeta = 9$ ps$^{-1}$.
Clearly this deterioration of transparency is due to the propagation
effects.
In addition we compare in Figs. {\ref{fig6}}(a) and (b)
 the population dynamics of states $| 1 \rangle$ and $| 3 \rangle$, $P_1(\zeta,\tau)$ and
$P_3(\zeta,\tau)$, at different optical depths $\mu_p \zeta =0$, $3$, $6$
and $9$ ps$^{-1}$.  All the parameters are chosen to be the same with
those for Fig. {\ref{fig4}}.
It is clear that a steady dark state is reached only at the entrance
 to the medium  ($\mu_p \zeta =0$) after irradiation
 with more than 60 pulses when both populations $P_1$ and $P_3$
(solid lines in Fig. {\ref{fig6}})  are time independent.

Next, we increase the coupling laser intensity so that $\Omega_{c0} =1$ THz
and hence the pulse area of the individual coupling laser pulse is
$\Omega_{c0}\tau_0 =1$, while keeping all other parameters exactly the same
with those for Fig. {\ref{fig4}}. The results are shown in  Fig. {\ref{fig7}}
for the first, $40^{th}$, and $120^{th}$ pulses in the probe pulse train.
Clearly, compared with the case of Fig. {\ref{fig4}},
the transmission of the probe laser pulses becomes much better
and  for the $120^{th}$ pulse we see an ideal EIT effect
where the probe laser pulse train  propagates without absorption.
The probe laser absorption for the  $120^{th}$ pulse  (not shown here)
takes negligible values  for optical depths  at least  up to $9$ ps$^{-1}$.
Similar to Fig. {\ref{fig5}} for $\Omega_{c0} =0.5$ THz,
we present in Fig. {\ref{fig8}} the dark and bright state populations
$P_D(\zeta,\tau)$ and $P_B(\zeta,\tau)$ for $\Omega_{c0} = 1$ THz
as a function of time at different optical depths.
We notice that the steady dark state population for $\Omega_{c0} = 1$ THz
is much closer to unity compared with the case of $\Omega_{c0} = 0.5$ THz,
 which indicates that even if the probe
and coupling laser pulses are short and form pulse trains, more ideal EIT
is realized by increasing the coupling laser intensity.
Clearly, after the irradiation of tens of pulses the $\Lambda$-type atomic medium
already reaches the steady dark state, and the temporal profile of the
probe pulse is hardly changed and the absorption of the probe pulses
is almost negligible at any optical depths.

Finally, in Figs. {\ref{fig9}}(a) and (b) we compare the population
dynamics of states $| 1 \rangle$ and $| 3 \rangle$, $P_1(\zeta,\tau)$ and
$P_3(\zeta,\tau)$, at different optical depths $\mu_p \zeta =0$, $3$, $6$
and $9$ ps$^{-1}$.
 All the parameters are chosen to be the same with those for Fig. {\ref{fig7}}.
A notable difference is seen between the results for $\Omega_{c0} =1$ THz
[Figs. {\ref{fig9}}(a) and (b)] and $0.5$ THz [Figs. {\ref{fig6}}(a) and (b)]:
 for $\Omega_{c0} =1$ THz the populations
$P_1$ and $P_3$ are time independent at different optical depths after
the irradiation with many  pulses (approximately 120),
 and they take almost the same values at any optical depth.
Therefore the population is trapped in a steady dark state,
 which is a superposition of the two lower states $| 1 \rangle$
and $| 3 \rangle$, and the upper state $| 2 \rangle$ remains practically
 unpopulated that give rise to propagation without absorption in Fig. {\ref{fig7}}.
For moderate coupling pulse area  $\Omega_{c0} \tau_0 =1$ the time scale to
 establish  the ideal EIT   and the generation
of a dark state is of the order of a few microseconds, while
for larger coupling pulse areas the EIT is established much
 faster in the nanoseconds regime.

\subsection{Propagation of  probe and coupling laser pulse trains  with
symmetric and asymmetric detunings}

Before closing this section we investigate how the laser detunings
influence the propagation dynamics of the probe laser pulse train.
Note that for all results presented until now in this paper, the central comb
teeth of both probe and coupling lasers are on exact resonance, i.e.,
$\delta_p=\delta_c=0$.
As mentioned before, the spectrum of a laser pulse train forms a frequency
comb with teeth which are separated by $2\pi/T$ and, in particular,
for a pulse train with a single pulse duration of $1$ ps and time interval of
$10$ ns the separation between the two successive comb teeth is $0.2 \pi$ GHz.

We consider two different choices of detunings for the probe and coupling
lasers as illustrated in Figs. {\ref{fig10}}(a) and (b),
which we call \textit{symmetric} and \textit{asymmetric detunings},
respectively.
For the symmetric detunings, all comb teeth of the probe and coupling laser
pulses match,
i.e., $\delta_p = \delta_c$, and the question is which comb teeth are on exact
resonance with the $| 1 \rangle$-$| 2 \rangle$ and $| 2 \rangle$-$| 3 \rangle$
transitions.
For the asymmetric detunings, the comb teeth of the probe and coupling lasers
at the opposite side with respect to the central peak in the frequency domain
match, i.e., $\delta_p = -\delta_c$.
Note that only one pair of the probe and coupling comb teeth is on exact
resonance with the corresponding transitions, as depicted in
Figs. {\ref{fig10}}(a) and (b), provided that the values
 of the probe and coupling detunings are
$\delta_{p} =  m_{p}  \;  \omega_r $ $ (\delta_{c} =   m_{c}\; \omega_r )$,
 with $m_{p}$ $ (m_c) $ an integer.
Therefore whatever the difference we might see between them in terms of probe
pulse propagation arises from the contributions of other comb teeth
which are off resonance.

In Fig. {\ref{fig11}} we compare the spatio-temporal changes
of the absolute value of probe laser field $|\Omega_p(\zeta,\tau)|$
 for the above choices of
detunings of the probe and coupling lasers:
(a) {\it symmetric detunings} $\delta_{p} =\delta_{c} \simeq 201.062$ GHz
(blue lines) and (b) {\it asymmetric detunings}
$\delta_{p} = -\delta_{c} \simeq 201.062$ GHz (red lines),
for the first (upper graph of Fig. {\ref{fig11}}),
10$^{th}$ (middle graph of Fig. {\ref{fig11}}), and 20$^{th}$
(lower graph of Fig. {\ref{fig11}}) pulses.
The symmetric (asymmetric) detunings  correspond to the $m_{p} = m_c = 320$
($m_{p} =- m_c = 320$) comb teeth of the probe and coupling lasers and
 we increase the coupling laser intensity such that $\Omega_{c0} = 2 $ THz.
All the rest of the parameters employed in
Fig. {\ref{fig11}} are the same with those for Fig. {\ref{fig4}}.
Clearly, we see some differences between the two cases,
 especially in the trailing edge of the probe pulse,
where the optical ringing \cite{Frohlich1991,Egorov2004}
occur for both cases, however the oscillations
 in the trailing edge disappear for symmetric detunings case
for larger $t/T$  because a  steady dark state is reached
quite fast after the interaction with the first $20$ pulses for
optical depths $\mu_p \zeta< 6$ ps$^{-1}$.
As mentioned before, these differences come from the contribution of all
other off-resonant comb modes of both lasers,
 Fig. {\ref{fig10}}(b) for the asymmetric case.

Accordingly, the dynamics of the dark and bright state populations
$P_D(\zeta,\tau)$ and $P_B(\zeta,\tau)$ is quite different in both cases,
as shown in Fig. {\ref{fig12}}.
For the symmetric detunings case the medium reaches a steady dark state
from the very beginning in terms of the number of pulses and optical depths
(blue lines in Fig. \ref{fig12}) and the dark and bright state populations
 slightly oscillate toward the establishment of EIT,
while for the asymmetric detunings the populations of both dark
and bright states exhibit a sew-saw pattern (red lines in Fig. \ref{fig12})
with envelopes that follow the profile of the dark and bright
state populations for symmetric detunings.
The oscillatory behavior of the dark and bright state populations
for asymmetric detunings is the result of the Rabi oscillations
of the probability amplitude of state $|3\rangle$, $c_3(\zeta,\tau)$,
that enters in the definitions of the dark and bright states
Eqs. (\ref{darkstate}) and (\ref{brightstate}).
The number of Rabi oscillations of $c_3(\zeta,\tau)$ during a single
pulse duration is proportional with the two-photon Raman detuning,
$\delta_p - \delta_c$, and depends on the comb modes order.
One of the conditions required to establish a dark state
 in case of a single pair of probe and coupling pulses is the two-photon
Raman resonance, but the EIT effect occurs over a small transparency frequency window,
 while for a train of laser pulses the best EIT effect occur, whenever
 all the probe and coupling comb teeth match ($\delta_p=\delta_c=  m_{p}\; \omega_r $),
 within a larger transparency window as shows  the population of the excited state
$P_2$, in Fig. \ref{fig13}, as a function of the probe laser detuning
at entrance to the medium (blue dashed lines for symmetric detunings).
The laser parameters in Fig. \ref{fig13} are the same as in Fig. \ref{fig11}
and the detuning dependence of the population of the upper excited state at
$\mu_p\zeta=0$ for a single pair of probe and coupling pulses ($N_{max}=1$)
 is compared to that of the probe and coupling laser pulse trains ($N_{max}=2,4,10$, and 20).
For the asymmetric detunings case, although the two-photon Raman detuning
  is not zero but multiple  of the repetition angular frequency,
$\delta_p-\delta_c= 2 \; m_{p}\; \omega_r $,
the atomic system also exhibits some degree of transparency
within a smaller transparency window
(red lines in Fig. \ref{fig13} for asymmetric detunings).
However it is clear from  Fig. \ref{fig13} that EIT is better established
for the case of symmetric detunings.

Of course there are many other choices for the probe and coupling laser
detunings, and the above choices are the two special cases we present.
Obviously, larger differences between pulse propagations with
symmetric and asymmetric detunings are obtained for energies
that do not  satisfy the one-photon resonance condition  \cite{Marian},
since the two-photon Raman resonance is always fulfilled
for symmetric detunings.
There is one exception, when a similar behavior of
 the pulse propagation leading to EIT is obtained for asymmetric detunings
 that do not correspond to one-photon resonance,
$\delta_p = -\delta_c = (2 \; m_{p}+1)\; \omega_r/2 $,
because the two-photon Raman detuning is again a multiple of
the repetition angular frequency and therefore the two-photon
 resonance condition is satisfied, as shows, in Fig. \ref{fig13}, the
 population of the excited state
for detunings that are odd multiples of $\omega_r/2 $.
In contrast with our findings for the $\Lambda$-type atom we expect that
the asymmetric detunings case should give the best EIT
 for a three level ladder-type atom
when the two-photon detuning dependence of the
 probability amplitude of state $|3\rangle$ in Eq. (\ref{amp3})
is given by $\delta_p + \delta_c$.

\section{Conclusions}
\label{su}

We have systematically studied the propagation of a pair of short laser pulse
trains in a three-level $\Lambda$-type atomic medium by simultaneously solving
the Maxwell-Schr\"{o}dinger equations.
First, for a single pair of probe and coupling pulses we have presented in Fig. \ref{fig2} the
 different spatio-temporal changes of the probe pulse with short duration (fs and ps)
compared to those with long duration (ns) for different coupling laser strengths.
For long pulse durations (ns) the oscillations of the probe pulse during the propagation
take place at the leading edge of the pulse and are induced
by the Rabi oscillations between the $| 1 \rangle$ and $| 3 \rangle$ levels.
The EIT effect for long pulse durations is established at large pulse areas
provided that $\Omega_{c0}>>\Omega_{p0}$.
In contrast, for short pulse durations (fs or ps) and small pulse areas the  trailing edge of the
 probe laser pulse is significantly distorted during the propagation
and the right wing of the pulse breaks up into several sub-pulses.
As the coupling pulse area increases the amplitude of the sub-pulses
 is reduced, the probe laser propagates undistorted  and an ideal EIT is established.
A second goal of this work is to compare,
\textit{for short pulses (ps)}, the propagation dynamics of a pair of
probe and coupling laser pulse trains  with a single pair of probe and
coupling laser pulses under the  conditions of EIT.
The main difference is that for a  single pair of probe and coupling lasers
the EIT is accomplished  for \textit{large coupling pulse areas}
($\Omega_{c0}\tau_0 >10^3$),
while for propagation of probe and coupling laser pulse trains the EIT
(as well the formation of a dark state), is reached for
\textit{moderate coupling pulse areas}
($\Omega_{c0}\tau_0 >1$),
and it strongly depends on the number of pulses in the train, as shown in Fig. \ref{fig7}.
We have discussed how the propagation dynamics of the laser pulse trains
through the medium under the EIT conditions could be modified by the
appropriate choice of laser parameters such as Rabi frequencies,
pulse durations, number of  pulses, and laser detunings.
Because of the use of short ps laser pulse trains and
moderate coupling field intensities, realization of
EIT has been found to be more demanding with ps train pulses compared with
the case of irradiation with a single pair of ns pulses. However, after the medium
interacts with a few tens of pulses at moderate coupling field intensities
EIT can be achieved.
We have shown that the attainment of EIT can be manipulated, depending on
which pairs of comb teeth are on resonance and a steady dark state can be
achieved whenever the two-photon detuning is an integer of the
 repetition angular frequency $\omega_r$.

\section{Acknowledgment}
This work was supported by a Grant-in-Aid for scientific research from the
Ministry of Education and Science of Japan.
The work by G.B. was partially supported by a research program Laplas 3 from
the National Authority for Scientific Research.
G.B. acknowledges the hospitality and the financial aid during her stay
at Kyoto University where a part of this work has been carried out.

\clearpage

\begin{figure}
\centering
\includegraphics[width=3.2in,angle=0]{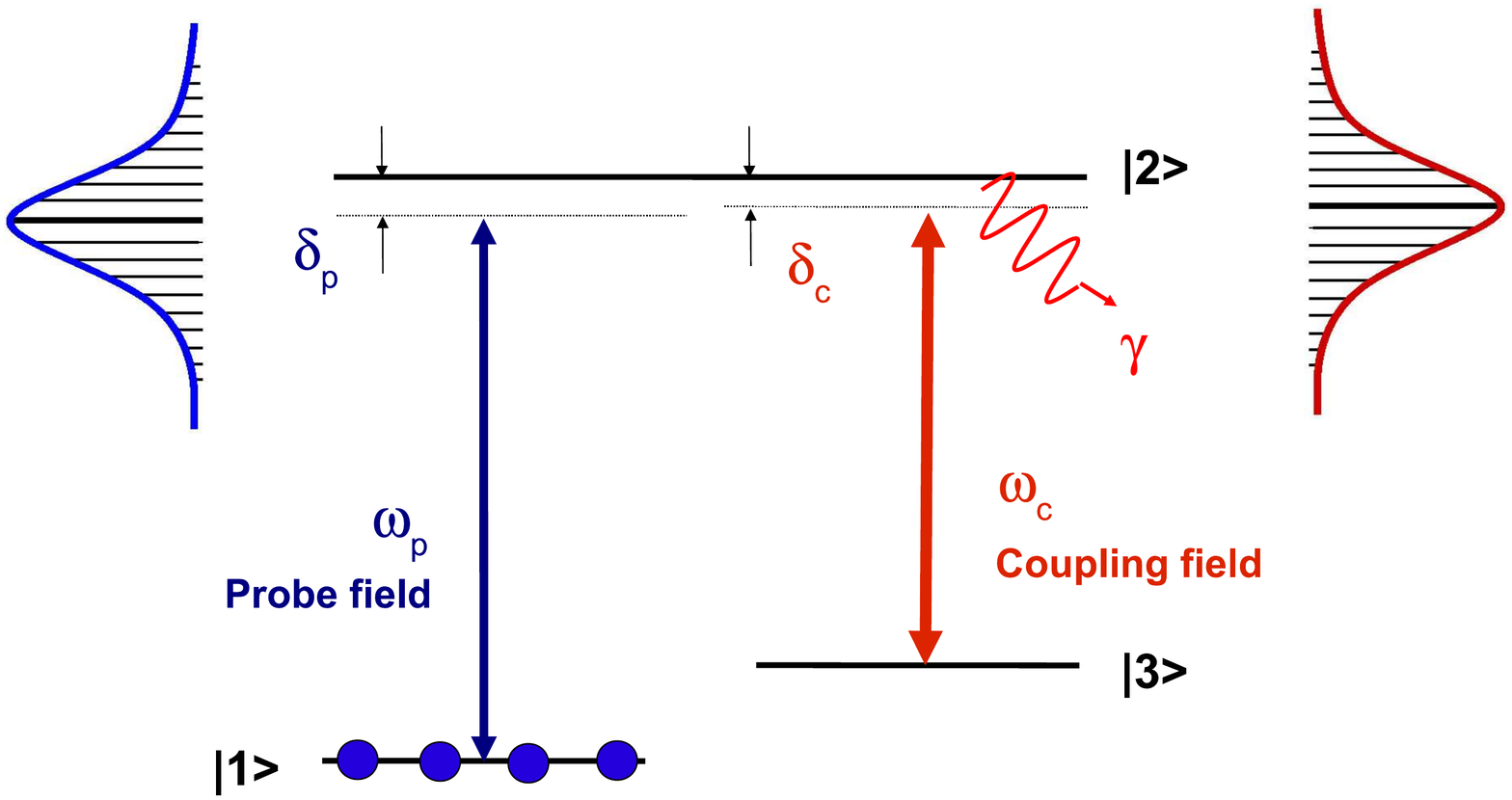}
\caption{
Level scheme for a three-level $\Lambda$-type atom interacting with the
probe and coupling lasers: States $| 1 \rangle$ and $| 2 \rangle$
are coupled by a probe laser with a photon energy of $\omega_p$, while
states $| 2 \rangle$ and $| 3 \rangle$ are coupled by a coupling laser
with a photon energy of $\omega_c$. If the laser is in a form of a pulse
train, the laser spectrum exhibits a frequency comb as illustrated.
}
\label{fig1}
\end{figure}

\begin{figure}
\vspace*{6mm}
\centering
\includegraphics[width=3.0in,angle=0]{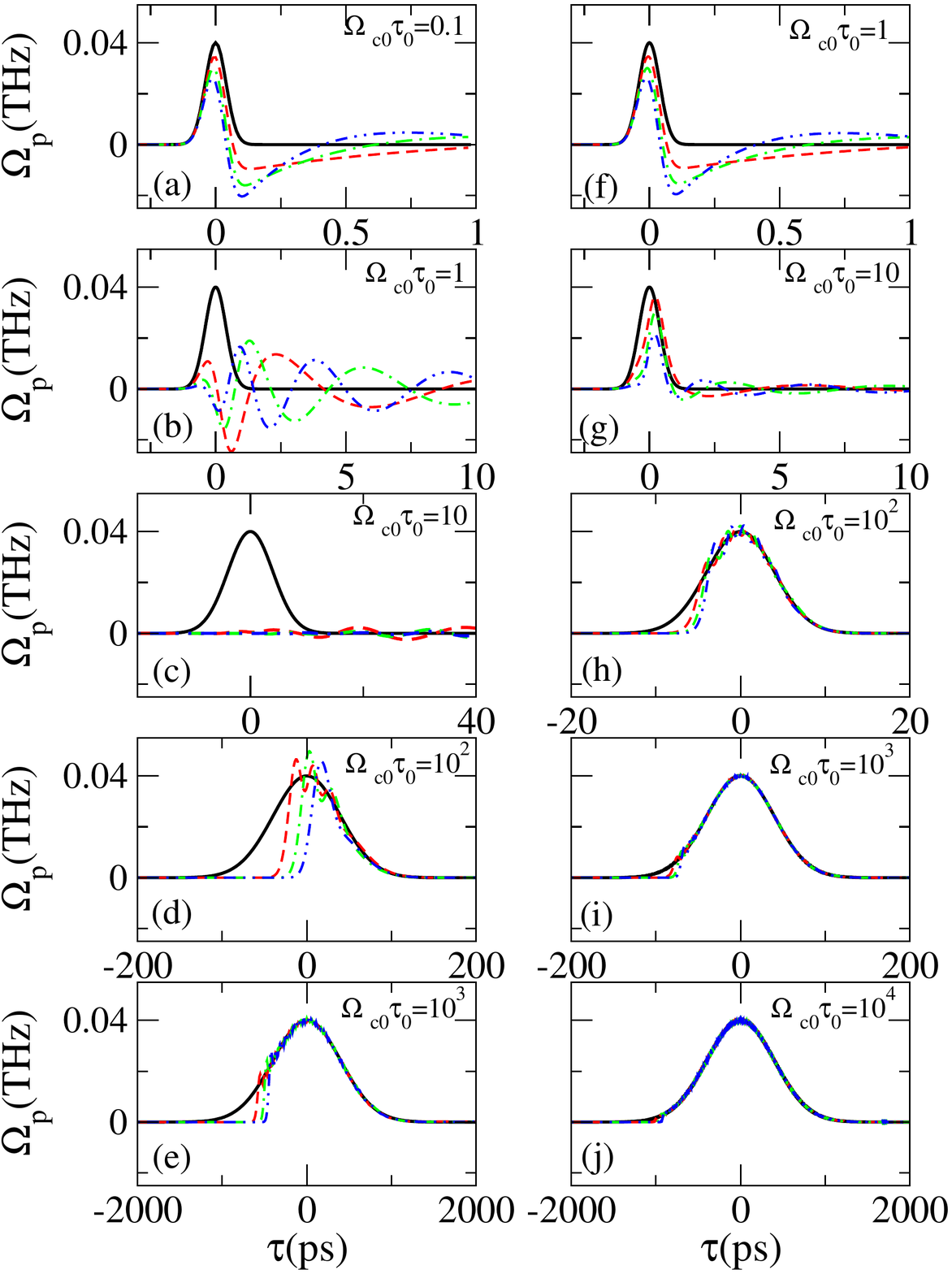}
\caption{
Temporal variation of the probe laser field $\Omega_p(\zeta,\tau)$
under the irradiation of a single pair of probe and coupling laser pulses
with the durations of $\tau_0=100$ fs [(a) and (f)],
1 ps [(b) and (g)], 10 ps [(c) and (h)], 100 ps [(d) and (i)], and
1 ns [(e) and (j)] at different optical depths $\mu_p \zeta= 0$ (black solid line),
$3 $ ps$^{-1}$ (red dashed line), $6 $ ps$^{-1}$ (green dot-dashed line), and
$9 $ ps$^{-1}$ (blue dot-dot-dashed line).
The employed parameters are $\Omega_{p0} = 0.04$ THz and $\Omega_{c0} =1$ THz
for graphs (a)-(e), and $\Omega_{p0} = 0.04$ THz and
$\Omega_{c0} = 10 $ THz for graphs (f)-(j).
For all graphs $\gamma=70 $ MHz and $\delta_p=\delta_c=0 $.
(For the interpretation of the references to colour in this figure legend,
the reader is referred to the web version of the paper.)
}
\label{fig2}
\end{figure}

\begin{figure}
\vspace*{6mm}
\centering
\includegraphics[width=3.in,angle=0]{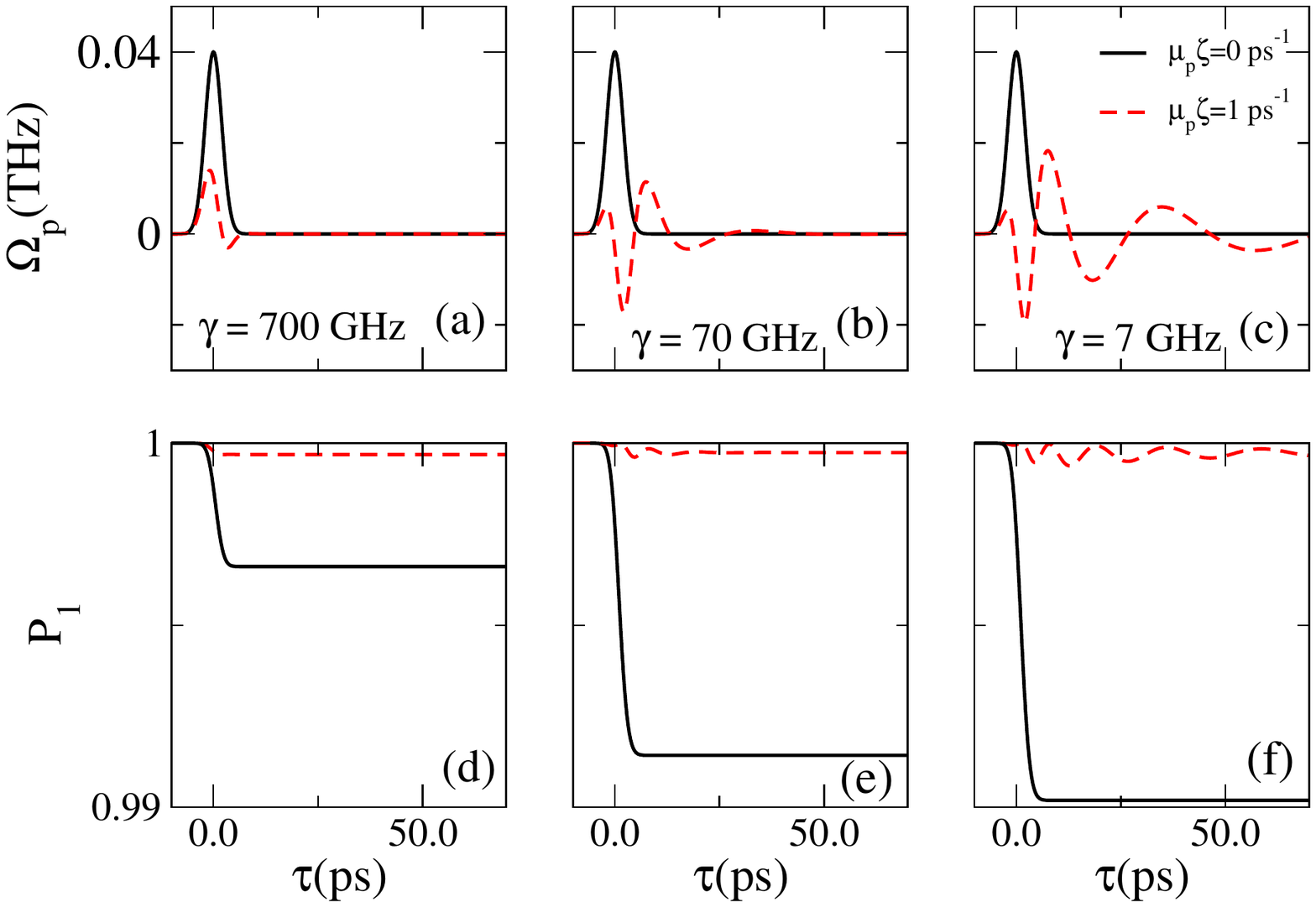}
\caption{
Temporal variation of the (a)-(c) probe laser field $\Omega_p(\zeta,\tau)$
and (d)-(f) ground state population $P_1(\zeta,\tau)$ at different
optical depths
$\mu_p \zeta= 0$ (solid line) and $1 $ ps$^{-1}$ (dashed line),
in the absence of the coupling laser field.
The employed parameters are $\Omega_{p0} =0.04$ THz,
$\delta_p=0$, $\tau_0 = 5$ ps, and
$\gamma=700$ GHz [(a) and (d)], or $\gamma=70$ GHz [(b) and (e)], or
$\gamma=7$ GHz [(c) and (f)], which result in the pulse area of
$\Omega_{p0}\tau_0 = 0.2$.
}
\label{fig3}
\end{figure}

\begin{figure}
\centering
\includegraphics[width=3.2in,angle=0]{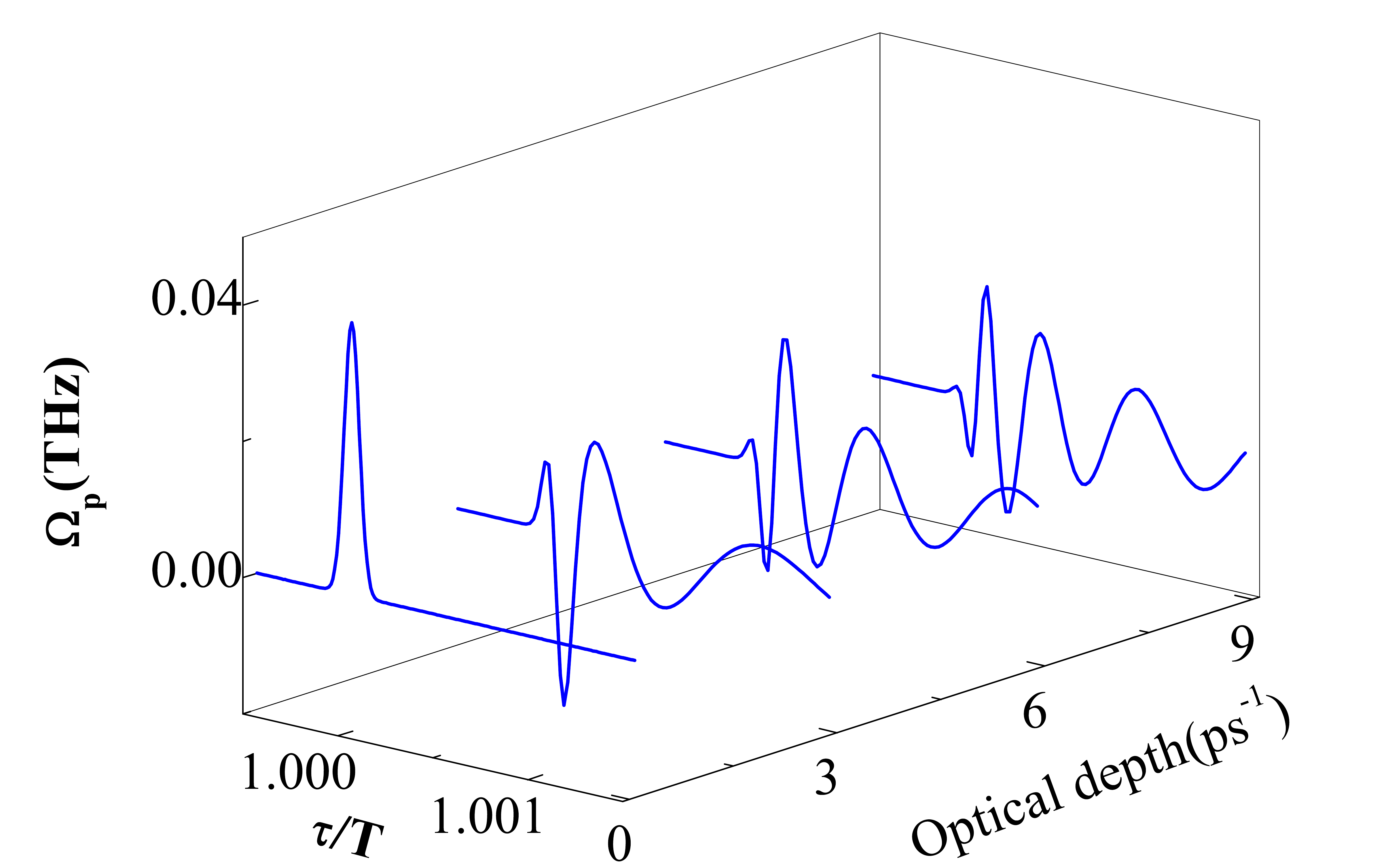}\\
\vspace*{0.2cm}
\includegraphics[width=3.2in,angle=0]{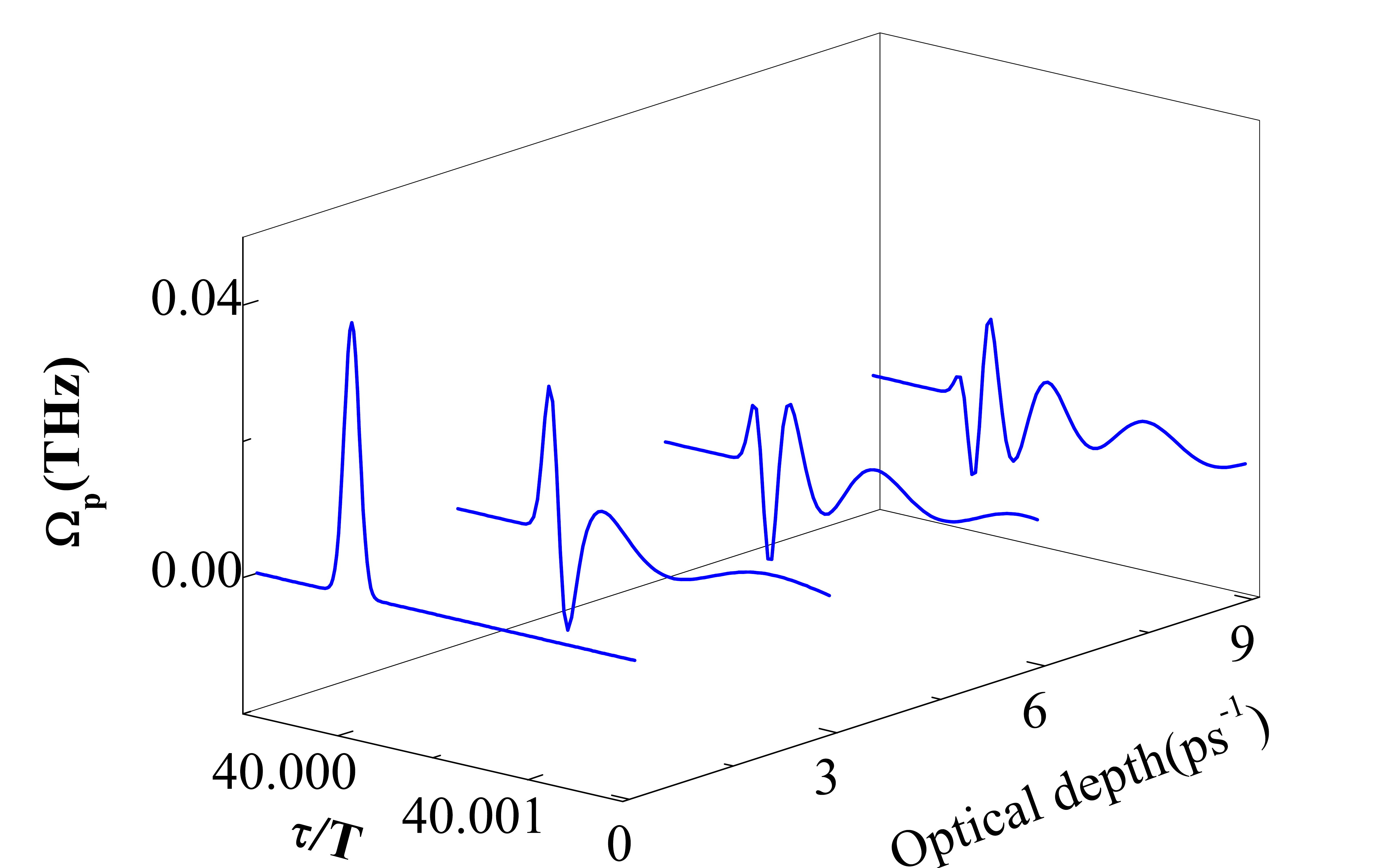}\\
\vspace*{0.2cm}
\includegraphics[width=3.2in,angle=0]{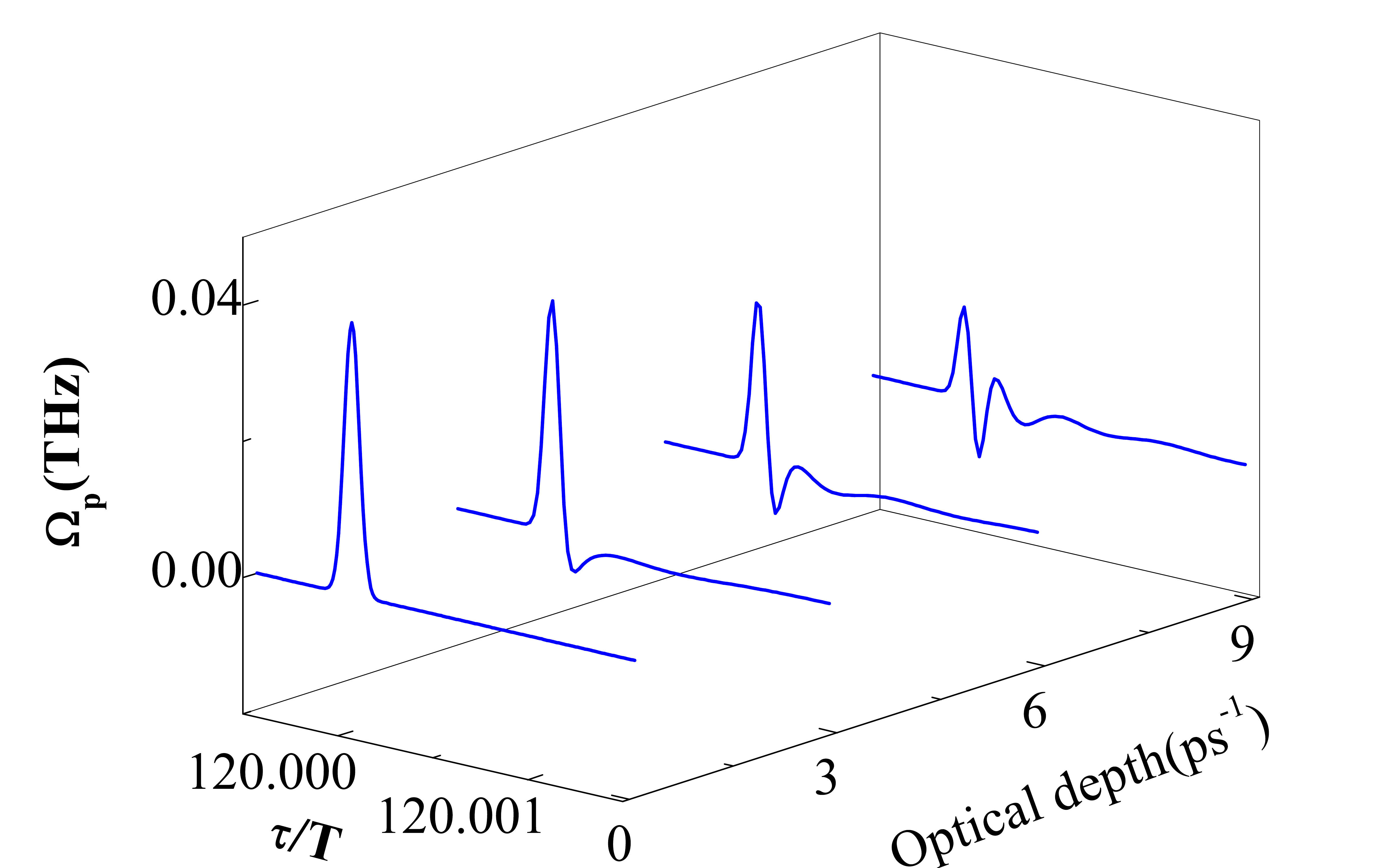}
\caption{
Spatio-temporal change of the probe laser field $\Omega_p(\zeta,\tau)$
under the presence of the coupling laser field at different optical depths
$\mu_p \zeta= 0$, $3$, $6$, and $9$ ps$^{-1}$.
The upper, middle, and lower figures are for the first, $40^{th}$, and
$120^{th}$ pulses in the probe pulse train.
The employed parameters are $\Omega_{p0} = 0.04$ THz,
$\Omega_{c0} = 0.5$ THz, $\delta_p=\delta_c=0$, $\tau_0 = 1$ ps,
$T=10$ ns, and $\gamma=70 $ MHz, which result in the pulse areas of
$\Omega_{p0}\tau_0 = 0.04$ and $\Omega_{c0}\tau_0 = 0.5$.
}
\label{fig4}
\end{figure}

\begin{figure}
\centering
\includegraphics[width=3.2in,angle=0]{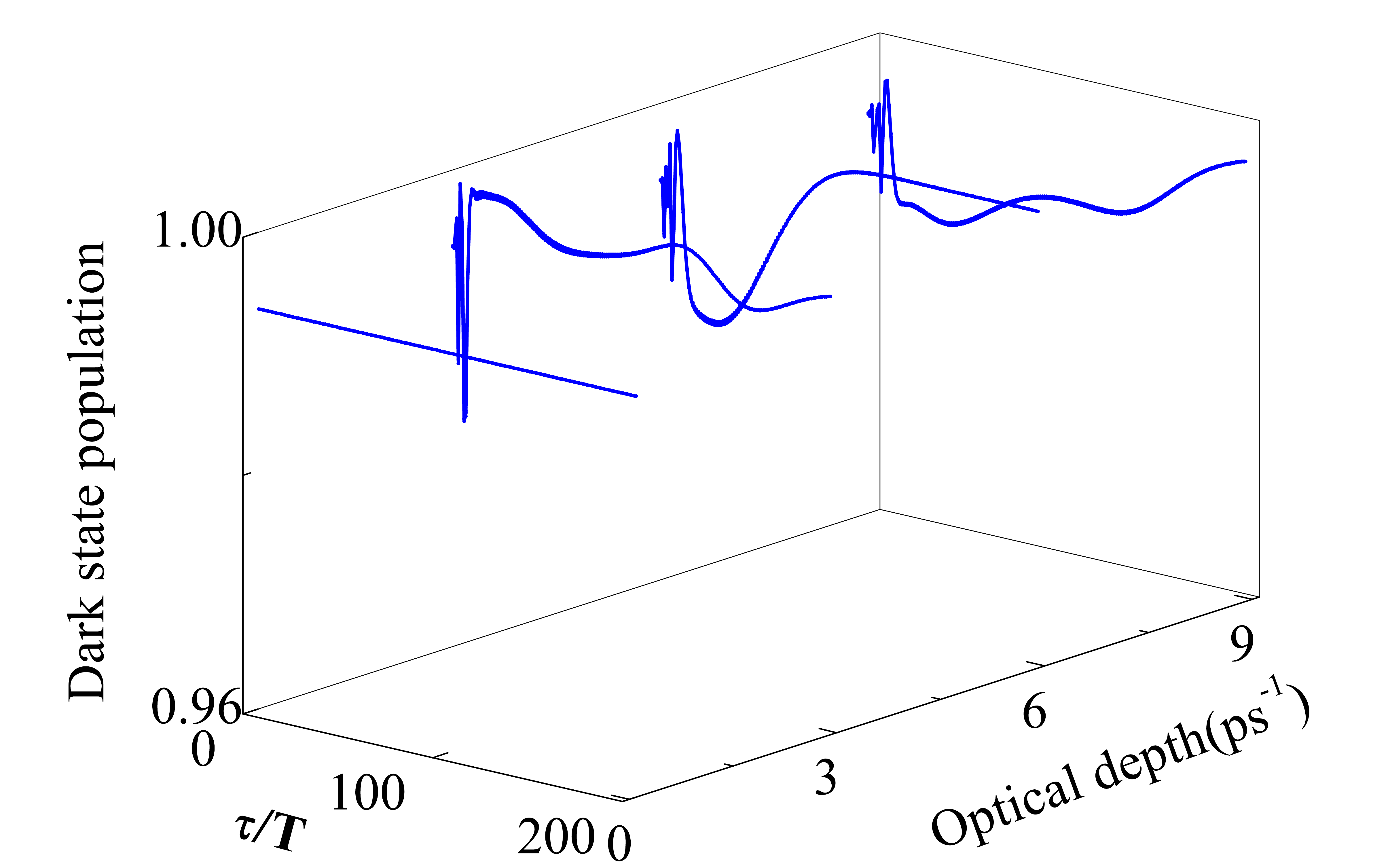}\\
\vspace*{0.2cm}
\includegraphics[width=3.2in,angle=0]{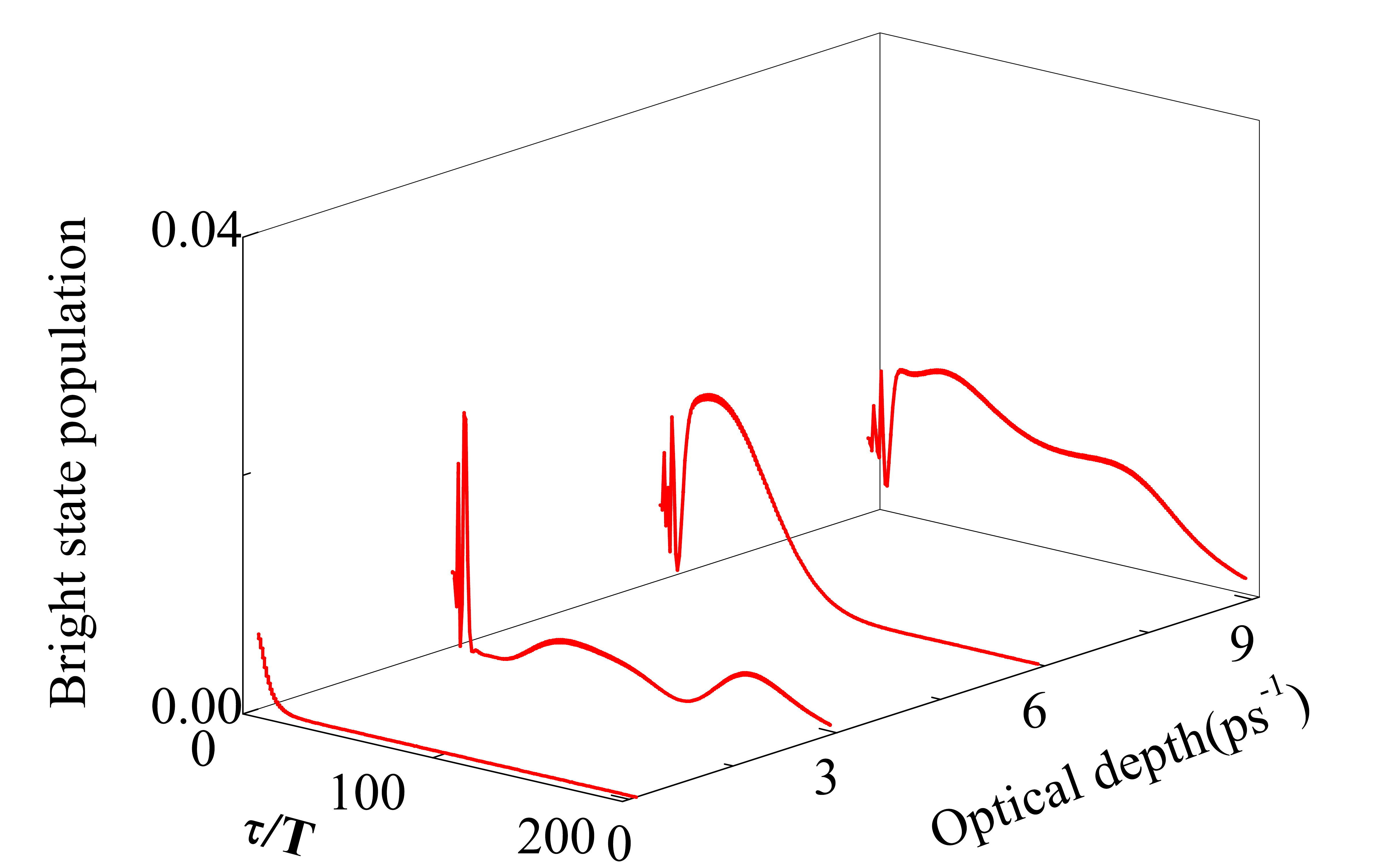}\\
\caption{
Spatio-temporal change of the dark and bright state populations
$P_D(\zeta,\tau)$ (upper figure) and $P_B(\zeta,\tau)$ (lower figure)
for different optical depths $\mu_p \zeta= 0$, $3$, $6$, and $9$ ps$^{-1}$.
All the parameters are the same with those for Fig. {\ref{fig4}}.
}
\label{fig5}
\end{figure}

\begin{figure}
\vspace{6mm}
\centering
\includegraphics[width=2.5in,angle=0]{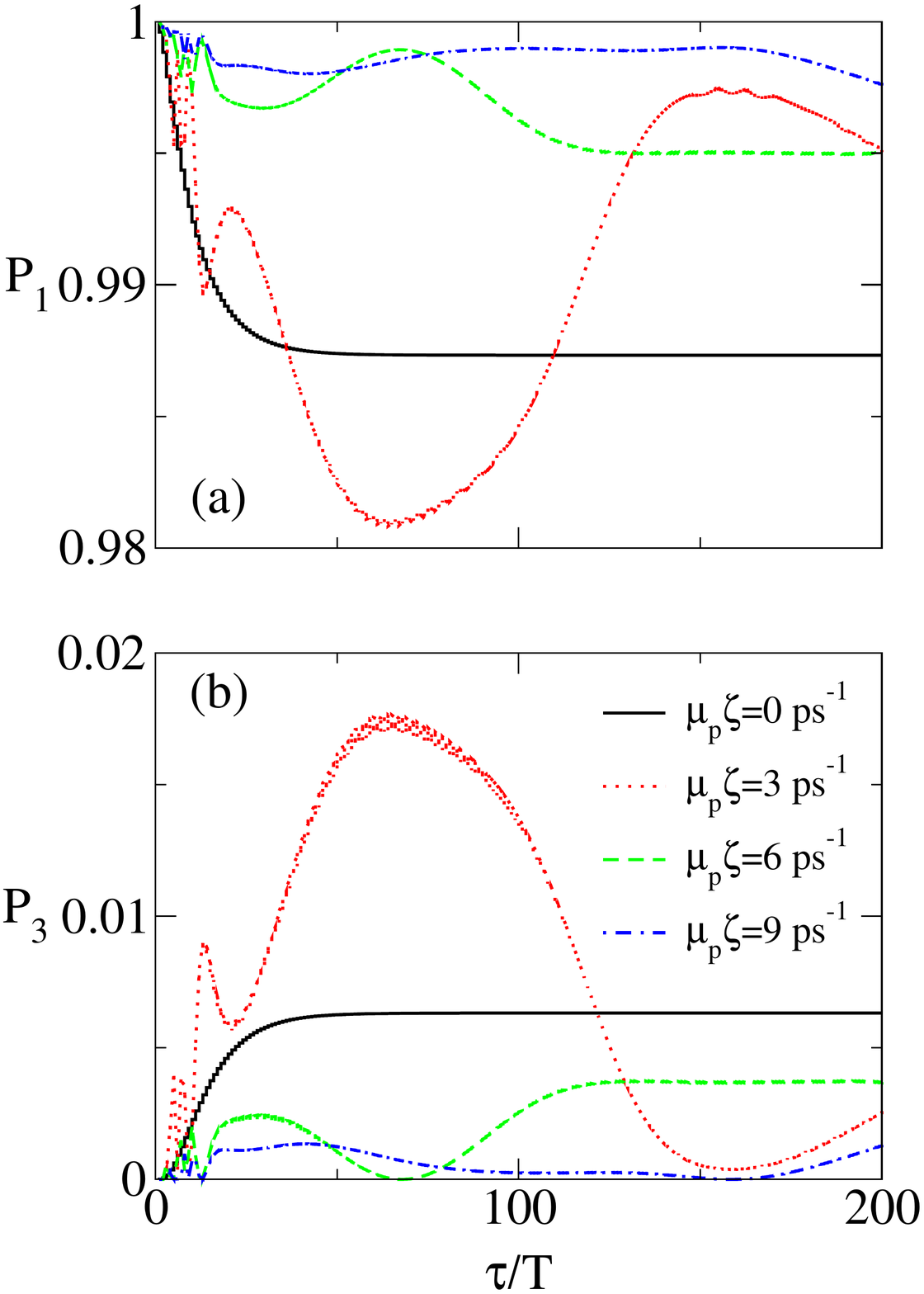}\\
\caption{
Ground  and excited state populations $P_1(\zeta,\tau)$ and $P_3(\zeta,\tau)$
as a function of time at different optical depths
 $\mu_p \zeta= 0$ (black solid line),
$3 $ ps$^{-1}$ (red dotted line), $6 $ ps$^{-1}$ (green dashed line), and
$9 $ ps$^{-1}$ (blue dot-dashed line).
The employed parameters are the same with those for Fig. {\ref{fig4}}.
(For the interpretation of the references to colour in this figure legend,
the reader is referred to the web version of the paper.)
}
\label{fig6}
\end{figure}

\begin{figure}
\centering
\includegraphics[width=3.2in,angle=0]{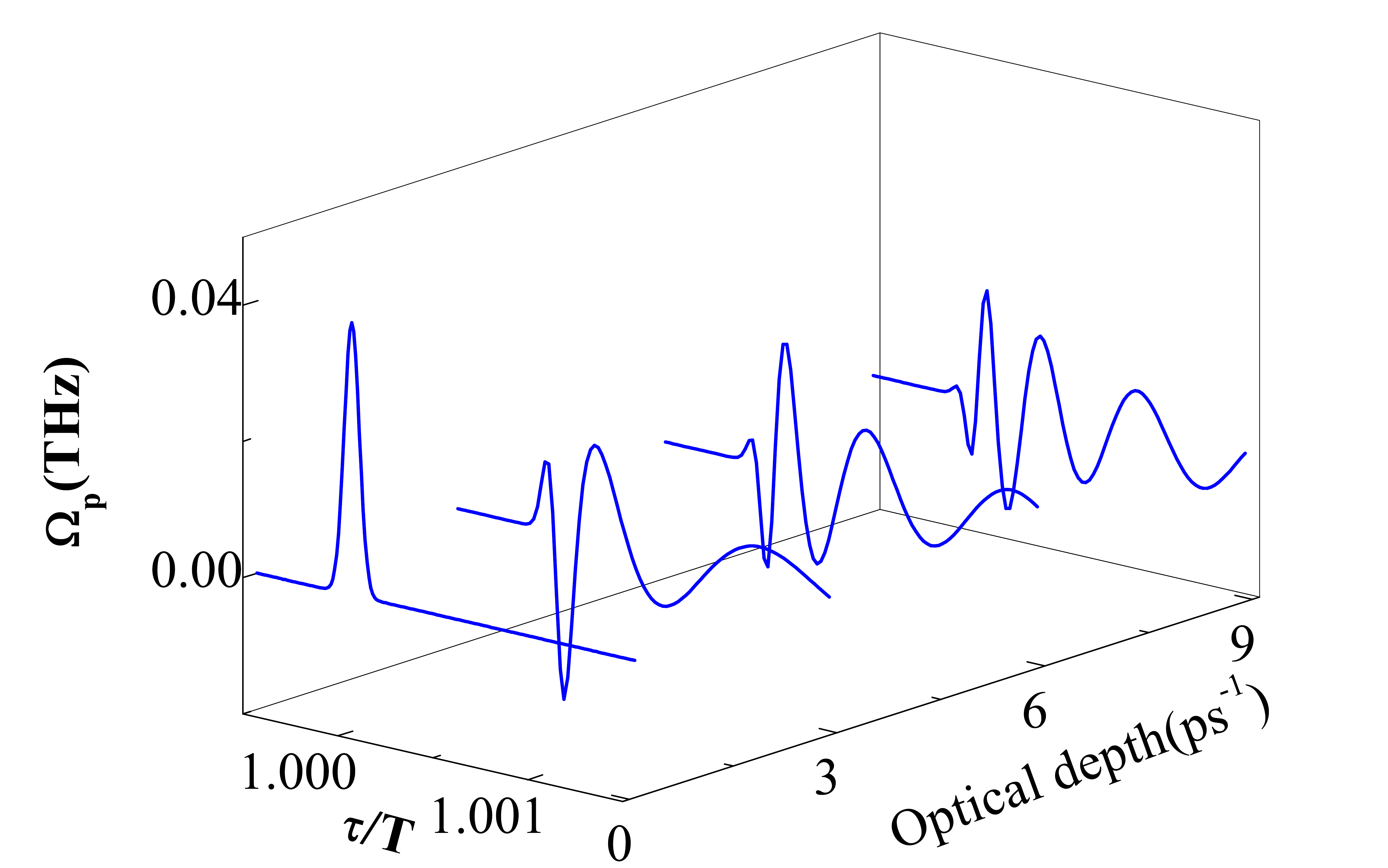}\\
\vspace*{0.2cm}
\includegraphics[width=3.2in,angle=0]{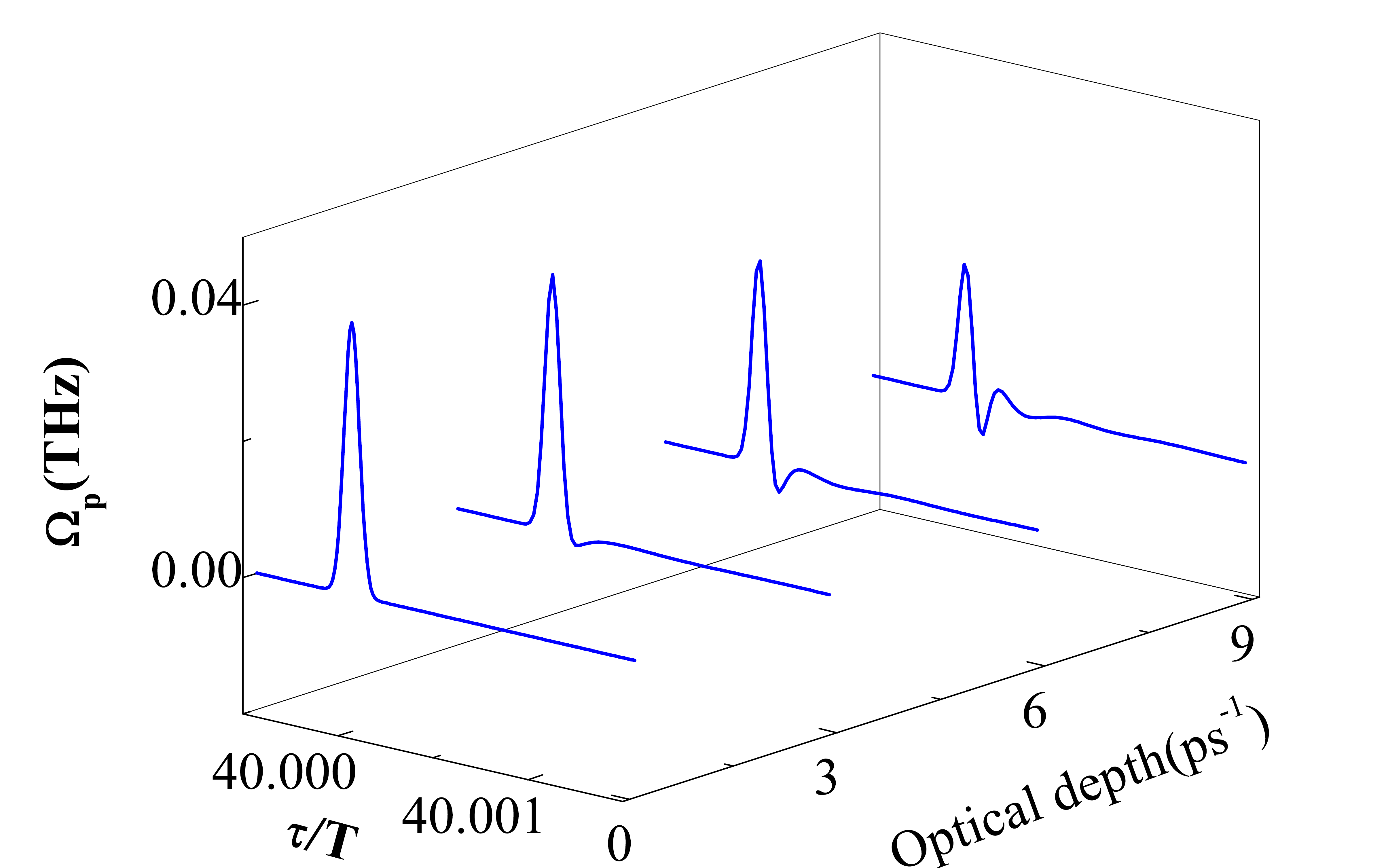}\\
\vspace*{0.2cm}
\includegraphics[width=3.2in,angle=0]{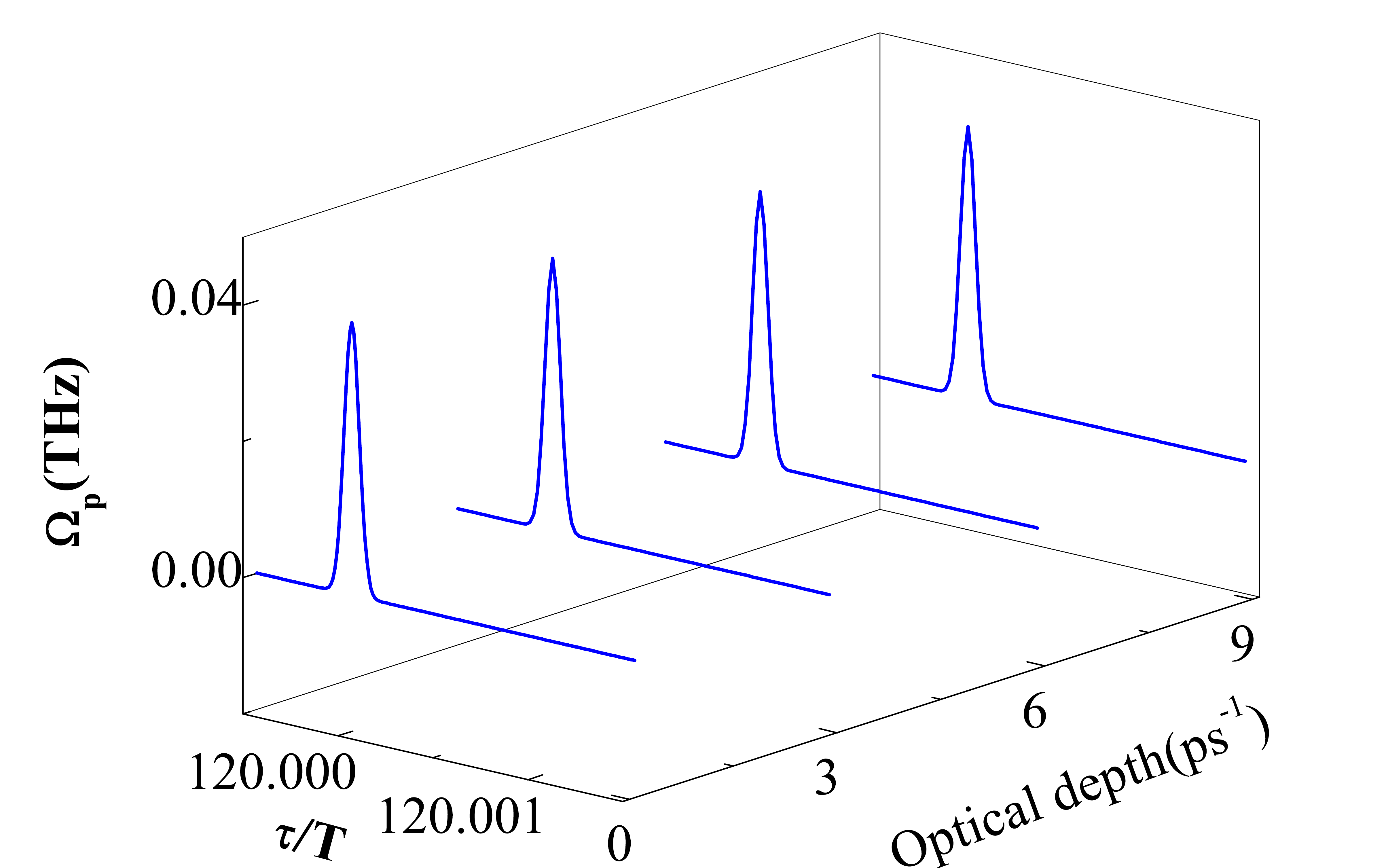}
\caption{
Similar to Fig. {\ref{fig4}} but with $\Omega_{c0} =1 $ THz, where the
rest of the parameters are the same with those for Fig. {\ref{fig4}}.
Accordingly the pulse areas are $\Omega_{p0}\tau_0$ = 0.04 and
$\Omega_{c0}\tau_0$ =1.
}
\label{fig7}
\end{figure}

\begin{figure}
\centering
\includegraphics[width=3.2in,angle=0]{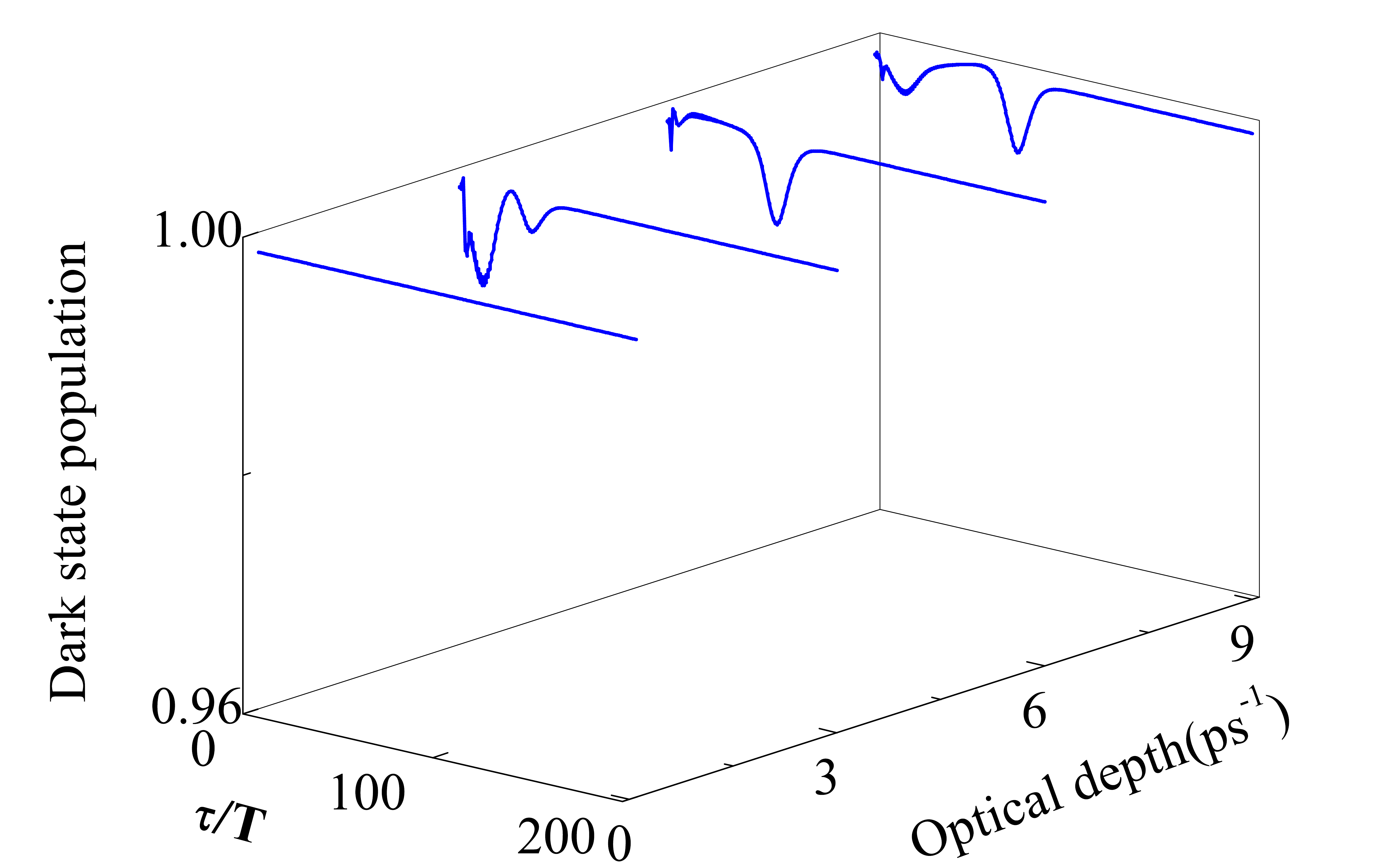}\\
\vspace*{0.2cm}
\includegraphics[width=3.2in,angle=0]{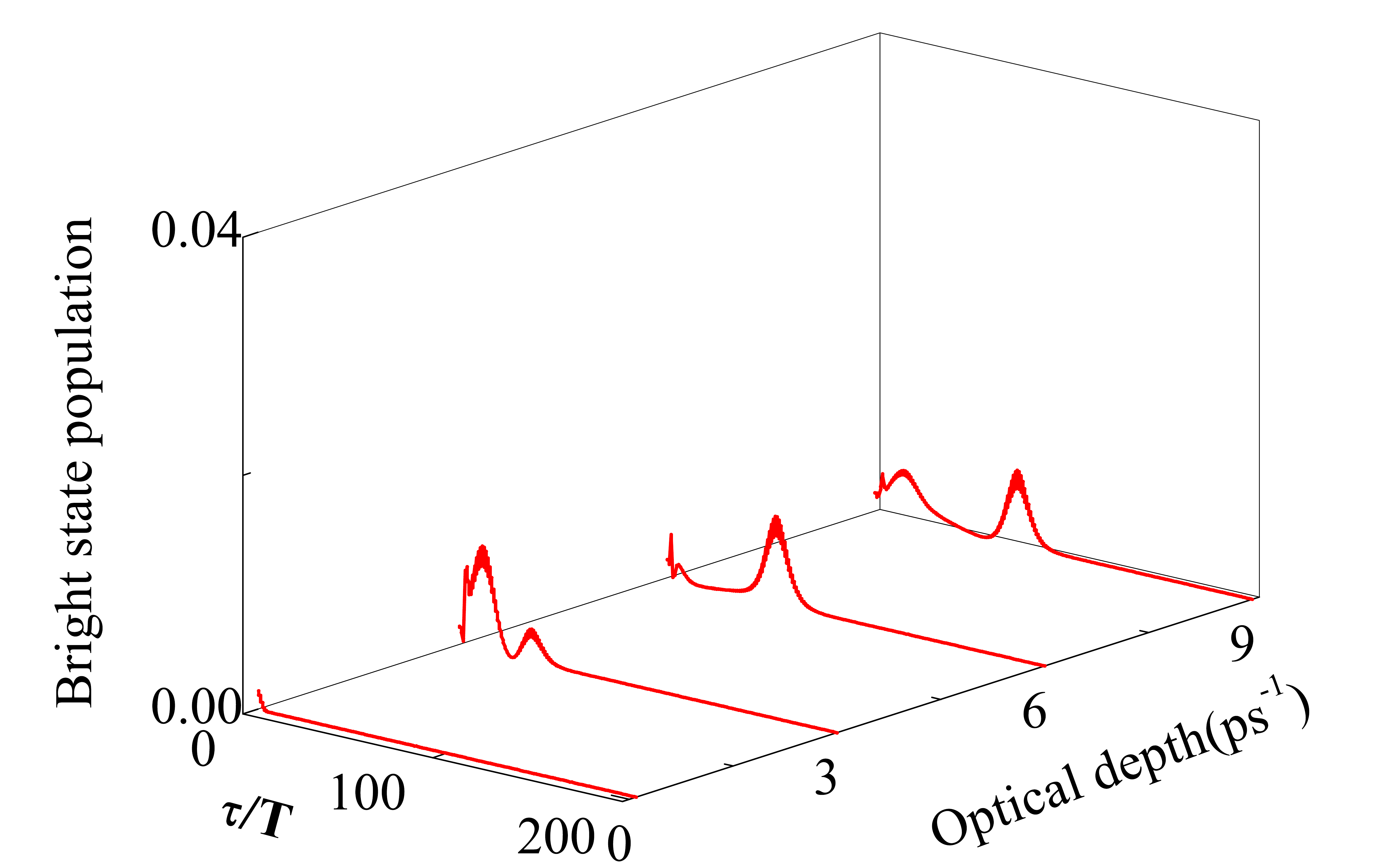}\\
\caption{
Spatio-temporal change of the dark and bright state populations
$P_D(\zeta,\tau)$ (upper figure) and $P_B(\zeta,\tau)$ (lower figure)
for different optical depths $\mu_p \zeta= 0$, $3$, $6$, and $9$ ps$^{-1}$.
All the parameters are the same with those for Fig. {\ref{fig7}}.
}
\label{fig8}
\end{figure}

\begin{figure}
\vspace{6mm}
\centering
\includegraphics[width=2.5in,angle=0]{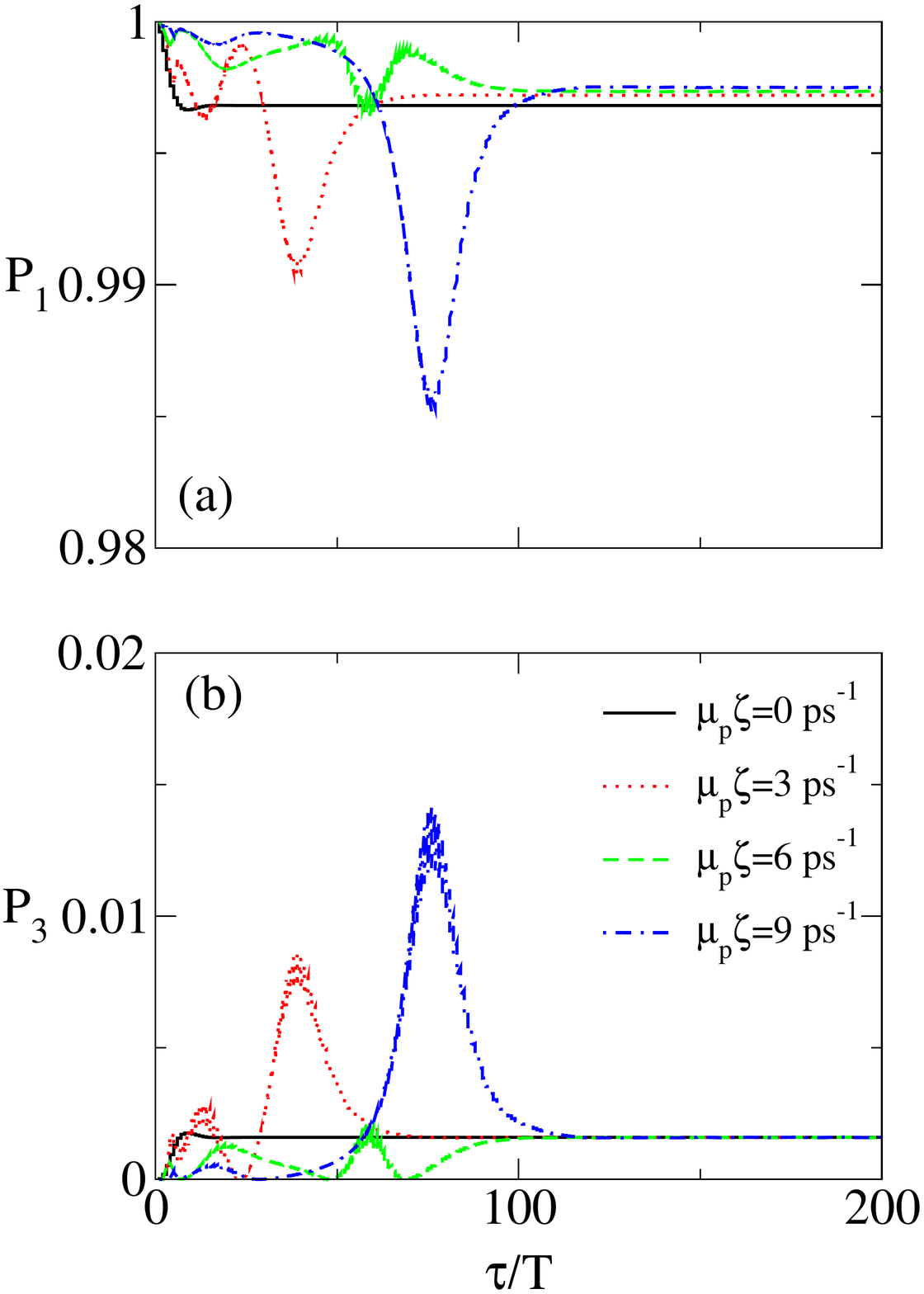}\\
\caption{
Ground  and excited state populations $P_1(\zeta,\tau)$ and $P_3(\zeta,\tau)$
as a function of time at different optical depths
 $\mu_p \zeta= 0$ (black solid line),
$3 $ ps$^{-1}$ (red dotted line), $6 $ ps$^{-1}$ (green dashed line), and
$9 $ ps$^{-1}$ (blue dot-dashed line).
The employed parameters are the same with those for Fig. {\ref{fig7}}.
(For the interpretation of the references to colour in this figure legend,
the reader is referred to the web version of the paper.)
}
\label{fig9}
\end{figure}

\begin{figure}
\centering
\includegraphics[width=3.25in,angle=0]{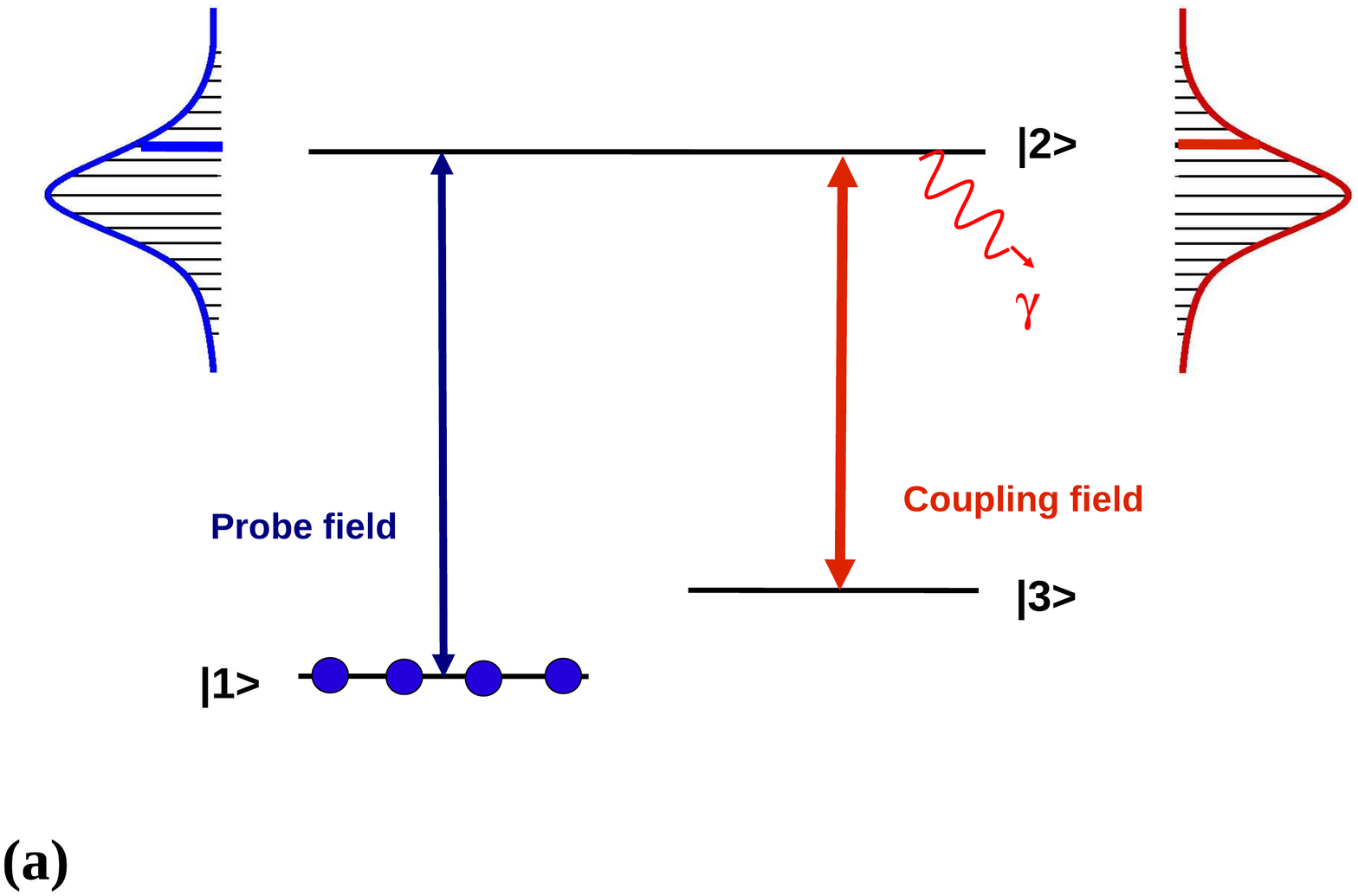}\\
\includegraphics[width=3.25in,angle=0]{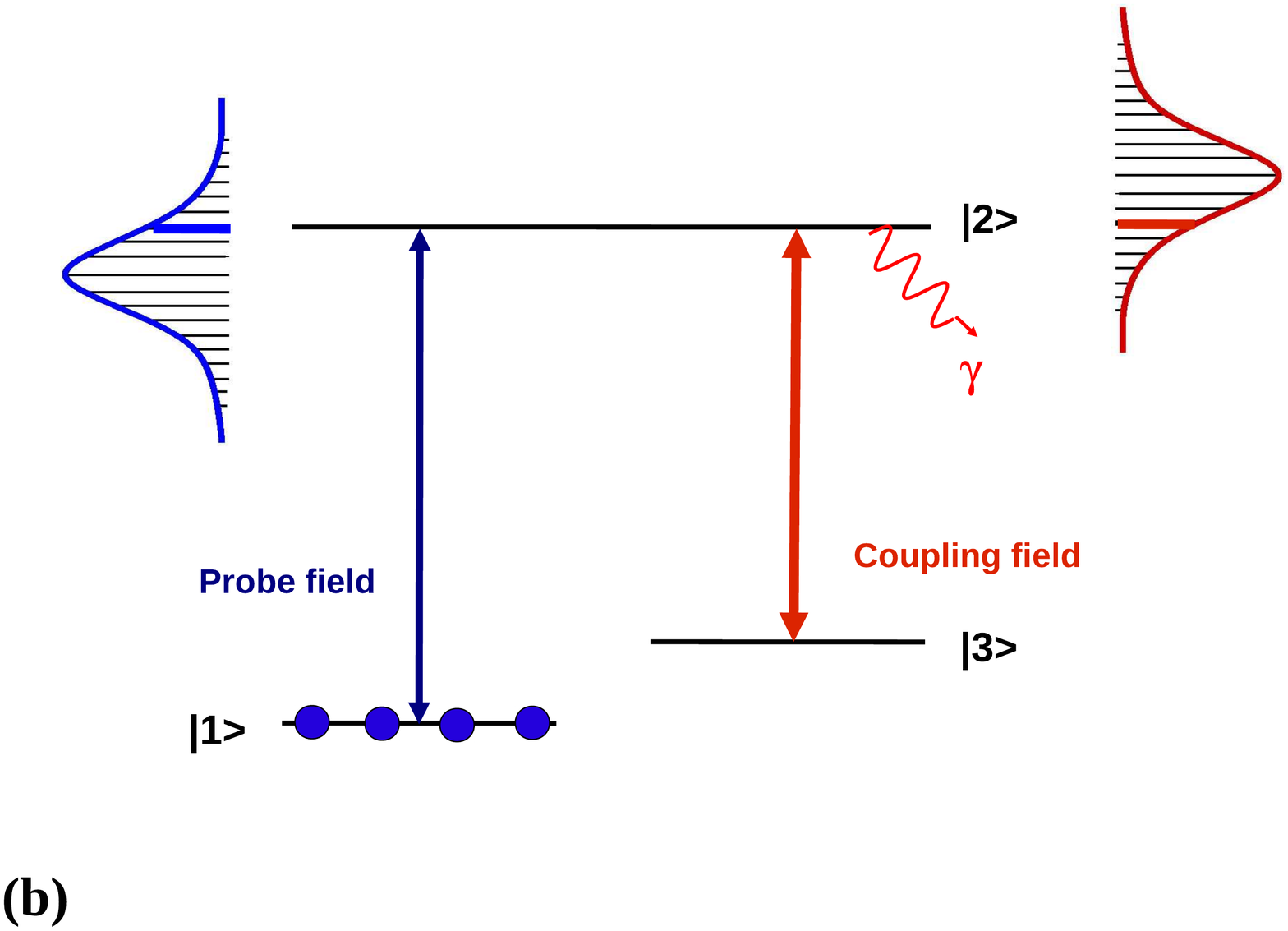}
\caption{(Color online)
Illustration of the (a) symmetric detunings $\delta_p=\delta_c$  and
(b) asymmetric detunings $\delta_p=-\delta_c$.
In either case one  of  probe as well as coupling laser comb tooth is
on exact resonance with the corresponding transition.
}
\label{fig10}
\end{figure}

\begin{figure}
\centering
\includegraphics[width=3.2in,angle=0]{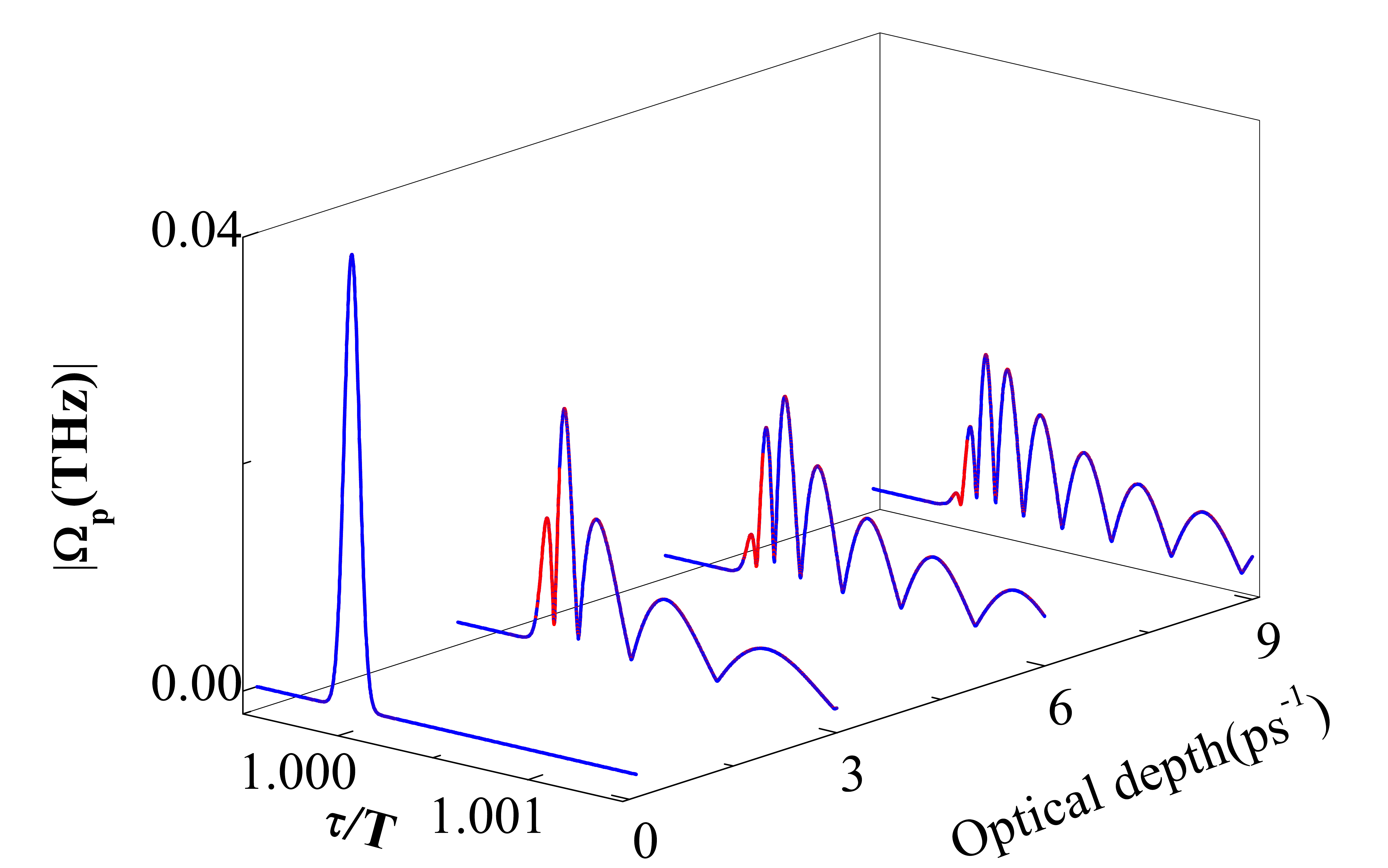}\\
\vspace*{0.2cm}
\includegraphics[width=3.2in,angle=0]{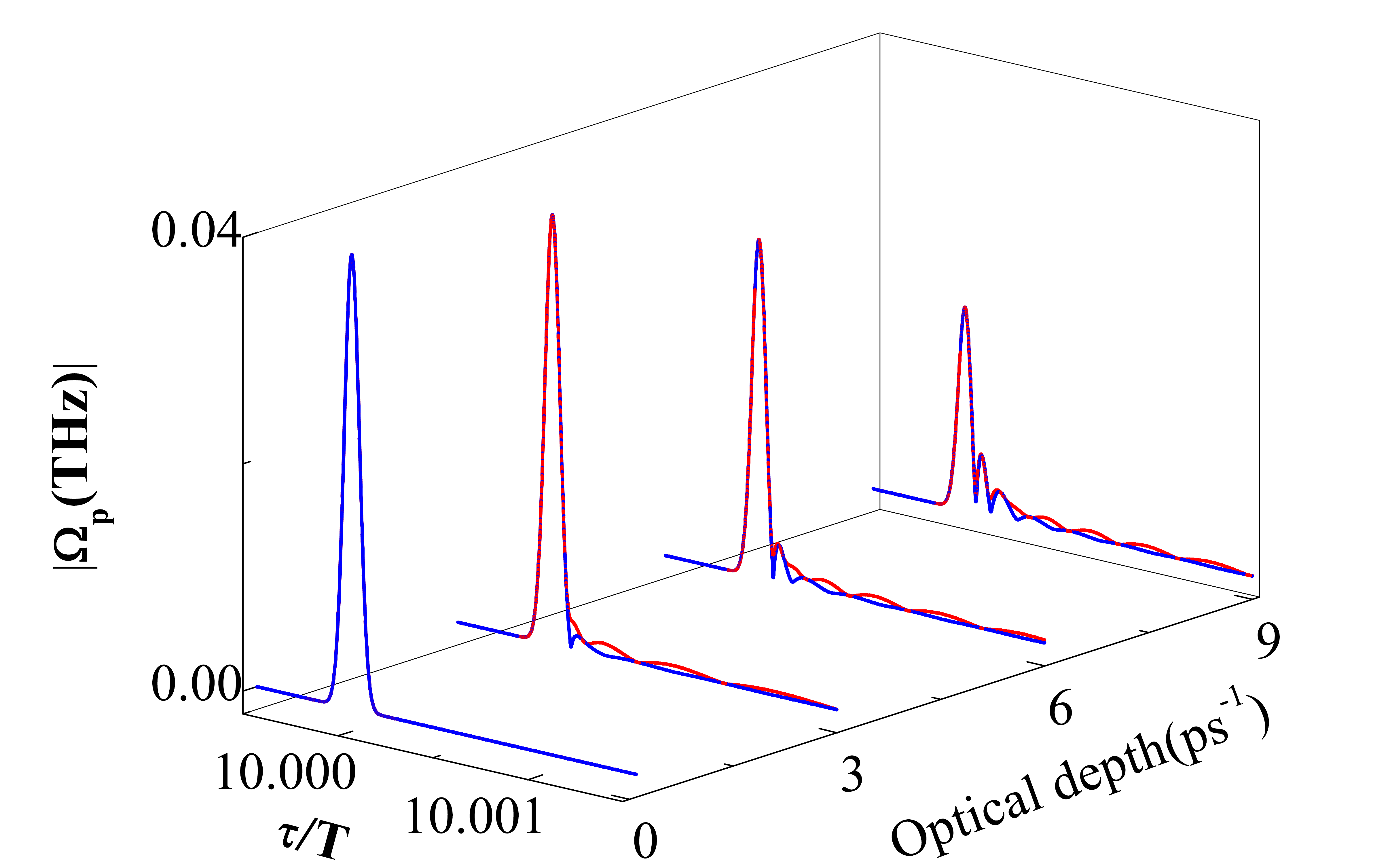}\\
\vspace*{0.2cm}
\includegraphics[width=3.2in,angle=0]{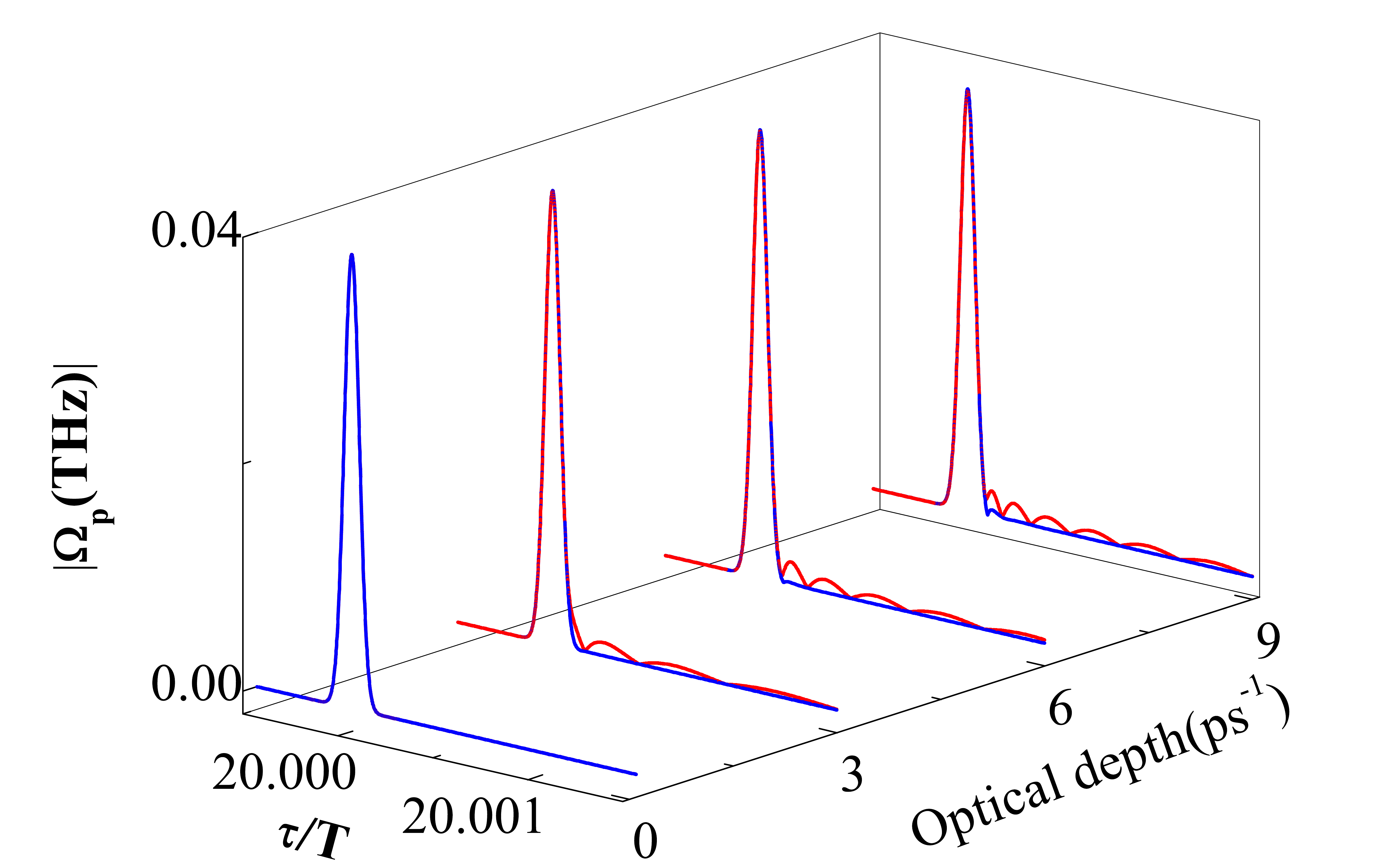}
\caption{
Spatio-temporal change of the probe laser field $|\Omega_p(\zeta,\tau)|$
at different optical depths $\mu_p \zeta= 0$, $3$, $6$, and $9$ ps$^{-1}$
under the symmetric (blue lines) and asymmetric detunings
(red lines).
The upper, middle, and lower figures show the first, $10^{th}$, and
$20^{th}$ pulses in the probe pulse train.
The choices of the symmetric and asymmetric detunings are
$\delta_{p} = \delta_{c} \simeq 201.062$ GHz and
$\delta_{p} =-\delta_{c} \simeq 201.062$ GHz, respectively.
$\Omega_{c0} =2 $ THz, the rest of the employed
 parameters are the same with those for
Fig. {\ref{fig4}} and accordingly the pulse areas are
$\Omega_{p0}\tau_0 = 0.04$ and $\Omega_{c0}\tau_0 =2$.
(For the interpretation of the references to colour in this figure legend,
the reader is referred to the web version of the paper.)
}
\label{fig11}
\end{figure}

\begin{figure}
\centering
\includegraphics[width=3.2in,angle=0]{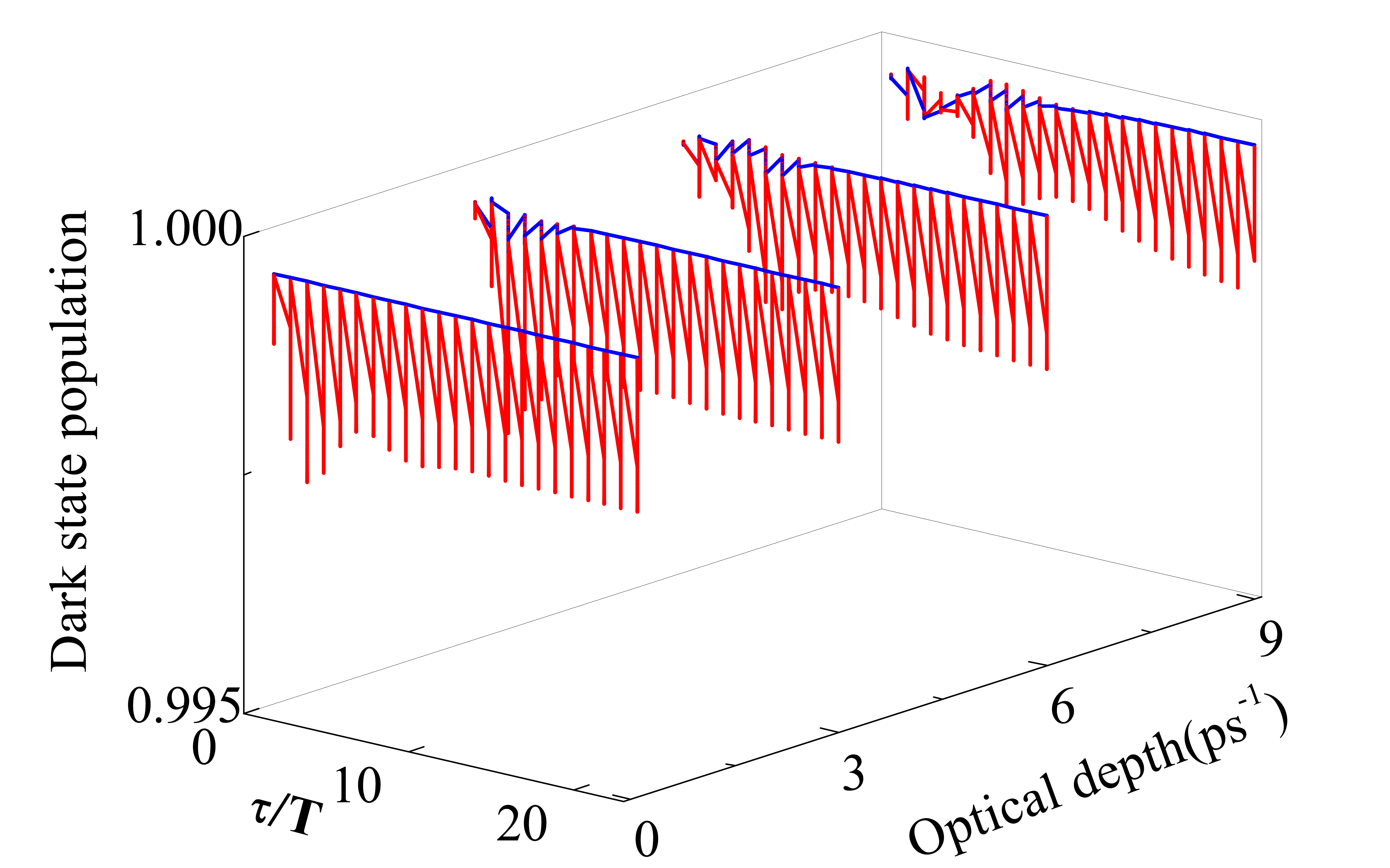}\\
\vspace*{0.2cm}
\includegraphics[width=3.2in,angle=0]{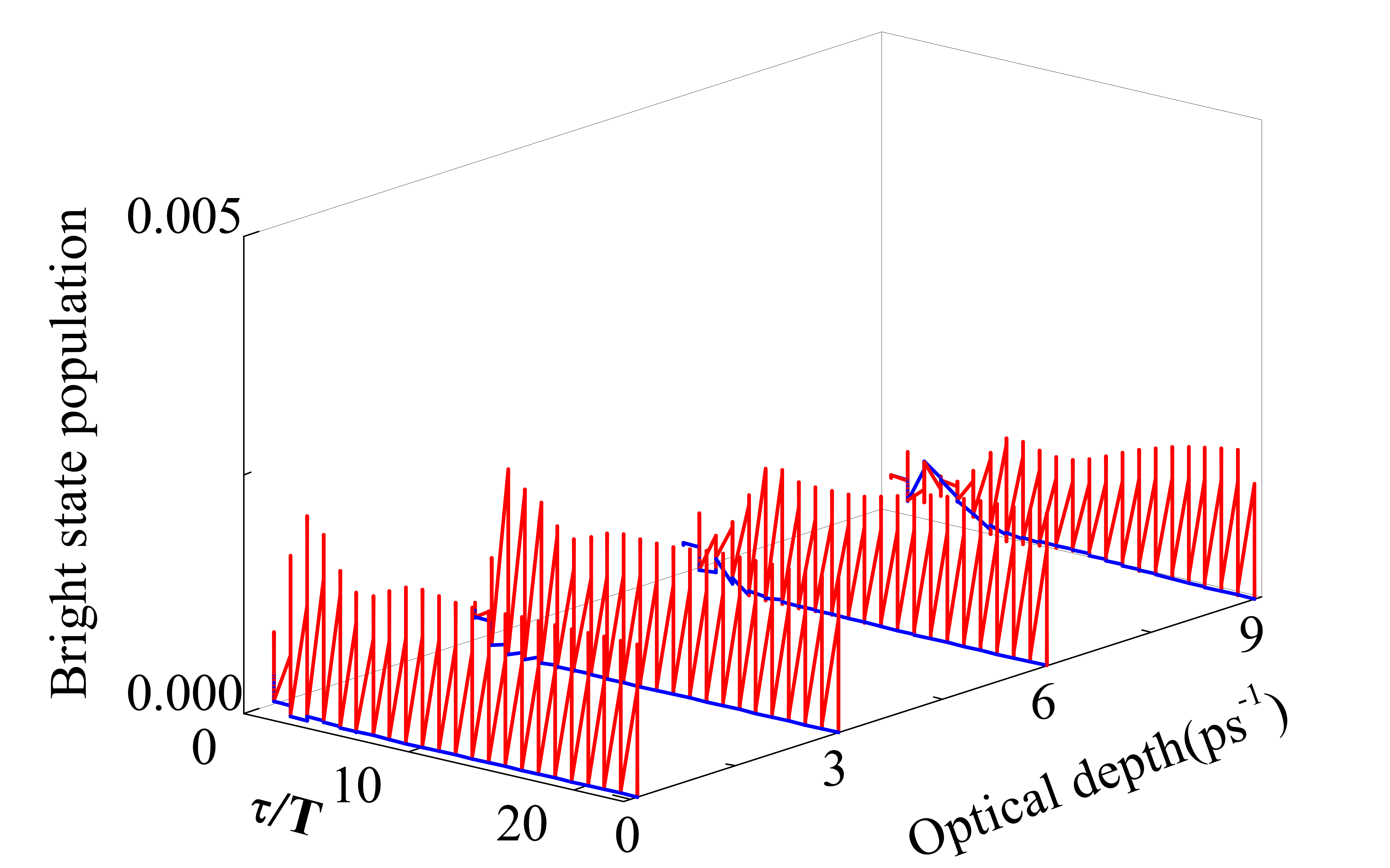}\\
\caption{
Spatio-temporal change of the dark and bright state populations
$P_D(\zeta,\tau)$ (upper figure) and $P_B(\zeta,\tau)$ (lower figure),
for different optical depths $\mu_p \zeta= 0$, $3$, $6$, and $9$ ps$^{-1}$,
under the symmetric (blue lines) and asymmetric detunings
(red lines, see-saw pattern).
All the parameters are the same with those for Fig. {\ref{fig11}}.
(For the interpretation of the references to colour in this figure legend,
the reader is referred to the web version of the paper.)
}
\label{fig12}
\end{figure}

\newpage
\begin{figure}
\vspace*{0.9cm}
\centering
\includegraphics[width=3in,angle=0]{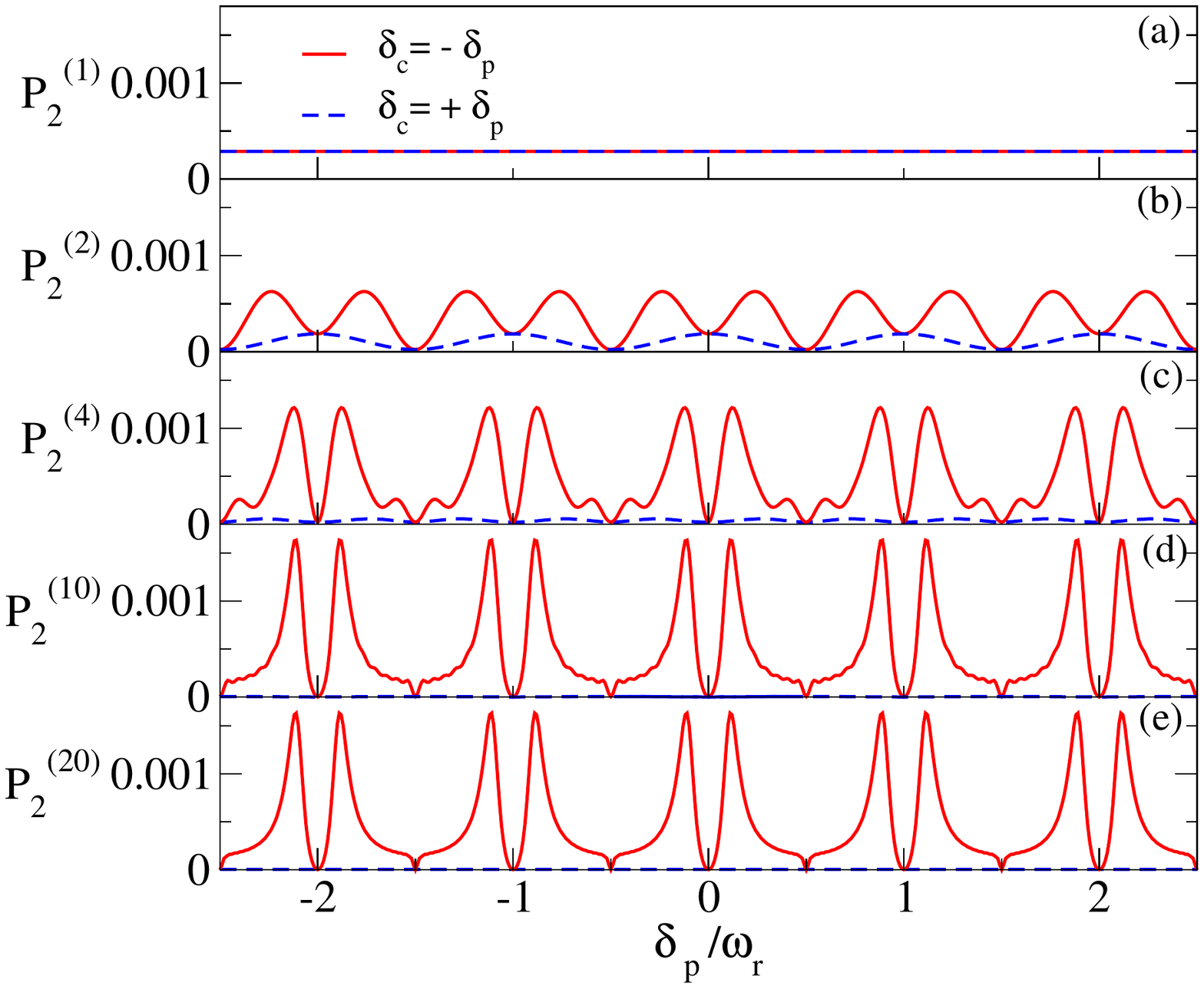}
\caption{
Population of the upper excited state $P_2$  as a function of the probe laser detuning,
at optical depth $\mu_p \zeta= 0$, under symmetric (blue dashed lines)
 and asymmetric detunings (red solid lines) for the first (a),
 $2^{nd}$ (b), $4^{th}$ (c), $10^{th}$ (d), and $20^{th}$ (e) probe pulse in the train.
All the parameters are the same with those for Fig. {\ref{fig11}}.
(For the interpretation of the references to colour in this figure legend,
the reader is referred to the web version of the paper.)
}
\label{fig13}
\end{figure}
\end{document}